\documentclass[preprint]{vgtc}               % preprint
\ifpdf%                                % if we use pdflatex
  \pdfoutput=1\relax                   % create PDFs from pdfLaTeX
  \usepackage{graphicx}                % allow us to embed graphics files
  \DeclareGraphicsExtensions{.pdf,.png,.jpg,.jpeg} % for pdflatex we expect .pdf, .png, or .jpg files
\else%                                 % else we use pure latex
  \usepackage{graphicx}                % allow us to embed graphics files
  \DeclareGraphicsExtensions{.eps}     % for pure latex we expect eps files
\fi%

\graphicspath{{figures/}{pictures/}{images/}{./}} % where to search for the images

\usepackage{microtype}                 % use micro-typography (slightly more compact, better to read)
\PassOptionsToPackage{warn}{textcomp}  % to address font issues with \textrightarrow
\usepackage{textcomp}                  % use better special symbols
\usepackage{mathptmx}                  % use matching math font
\usepackage{times}                     % we use Times as the main font
\usepackage{cite}                      % needed to automatically sort the references
\usepackage{tabu}                      % only used for the table example
\usepackage{booktabs}                  % only used for the table example
\usepackage{multirow}
\usepackage{subfig}
\usepackage{comment}
\usepackage{transparent}
\usepackage{amssymb}
\usepackage{environ}
\usepackage{csvsimple,etoolbox}
\usepackage{tikz}
\usetikzlibrary{arrows,arrows.meta,patterns,calc,shapes,decorations.pathreplacing,matrix,backgrounds,positioning}
\usepackage{tcolorbox}
\usepackage{pgfplots}
\pgfplotsset{compat = newest}
\usepgfplotslibrary{patchplots}
\usepackage{adjustbox}
\usepackage{amsmath}
\usepackage{xspace}
\usepackage[version=4]{mhchem}
\usepackage{appendix}
\usepackage{array}

\DeclareMathAlphabet{\mathcal}{OMS}{cmsy}{m}{n}

\usepackage{amsmath}				% Math Library
\usepackage{amssymb}
\newcommand{\R}{\mathbb{R}}

\newcommand{\etal}{\textit{et al.}\xspace}

\newcommand{\domain}{\mathcal{K}}
\newcommand{\dMan}{\mathcal{D}}
\newcommand{\vertices}{\mathcal{V}}

\newcommand{\sublevelset}[1]{{#1}^{-1}_{-\infty}}

\newcommand{\Cforward}{\mathcal{C}^+}
\newcommand{\Cbackward}{\mathcal{C}^-}
\newcommand{\Overlap}{\mathcal{O}}
\newcommand{\FOverlap}{\mathcal{FO}}
\newcommand{\FCmatrix}{\mathcal{FC}}

\newcommand{\Cmatrix}{\mathcal{C}}
\newcommand{\aMan}{\mathcal{A}}
\newcommand{\bMan}{\mathcal{B}}

\definecolor{ilineColor}{HTML}{000000}
\definecolor{id0}{HTML}{fb9a99}
\definecolor{id1}{HTML}{33a02c}
\definecolor{id0overlapid1}{HTML}{84906b}
\definecolor{hydrogen}{rgb}{1, 1, 1}
\definecolor{oxygen}{rgb}{1, 0.0509804, 0.0509804}
\definecolor{carbon}{rgb}{0.564706, 0.564706, 0.564706}
\definecolor{changed}{rgb}{0, 0, 0}

\tikzstyle{arrow} = [-{Stealth[scale=0.8]}]
\tikzstyle{iline} = [line width=0.1cm, densely dashed, ilineColor, arrow]
\tikzstyle{iNode} = [black,fill=black,stroke=black,scale=1.5]
\tikzstyle{iSeed} = [iNode,circle,inner sep=1pt]
\tikzstyle{catA} = [iNode,circle,inner sep=0.5pt, fill=id0]
\tikzstyle{catB} = [iNode,circle,inner sep=0.5pt, fill=id1]
\tikzstyle{catA-man} = [iNode,inner sep=0.5pt, fill=id0]
\tikzstyle{catB-man} = [iNode,inner sep=0.5pt, fill=id1]
\tikzstyle{catOverlap} = [iNode,inner sep=0.5pt, fill=id0overlapid1]
\tikzstyle{iSample} = [fill=white,circle,inner sep=1pt, scale=2]
\tikzstyle{sampleLine} = [line width = 1pt, dashed, white]
\tikzstyle{sampleCircle} = [circle,fill=white, opacity=0.25]
\tikzstyle{matrixCell} = [rectangle,draw,minimum width=3em,minimum height=1.6em,font=\ttfamily,anchor=south]

\newcommand{\changed}[1]{\textcolor{black}{#1}}
\onlineid{1012}

\vgtccategory{Research}

\vgtcinsertpkg

\title{Probabilistic Gradient-Based Extrema Tracking}

\author{Emma Nilsson\thanks{e-mail: \{emma.nilsson$|$talha.bin.masood$|$ingrid.hotz\}@liu.se}\\ %
\and Jonas Lukasczyk\thanks{e-mail: \{lukasczyk$|$garth\}@rptu.de}\\ 
\and Talha Bin Masood\footnotemark[1]\\ %
\and Christoph Garth\footnotemark[2]
\and Ingrid Hotz\footnotemark[1]\\ %
}
\affiliation{\scriptsize Linköping University\footnotemark[1] \\ University of Kaiserslautern-Landau\footnotemark[2]}

\teaser{
  \centering
    \adjustbox{width=\linewidth}{
        \begin{tikzpicture}[scale=1]
    \foreach \x [evaluate=\x as \xpos using (4.72-(-4.72)) * (\x - 40)/30 +(-4.72)] in {40, 45,  ..., 70}{
        \draw[gray, line width=0.05mm] (\xpos, -1.5) -- (\xpos, 1.5);
        
        \node[scale=1] at (\xpos, -1.75) {\tiny \x};
    }

    \node(graph) at (0,0) {
        \includegraphics[height=3cm, trim={4cm 5cm 4cm 18cm},clip]{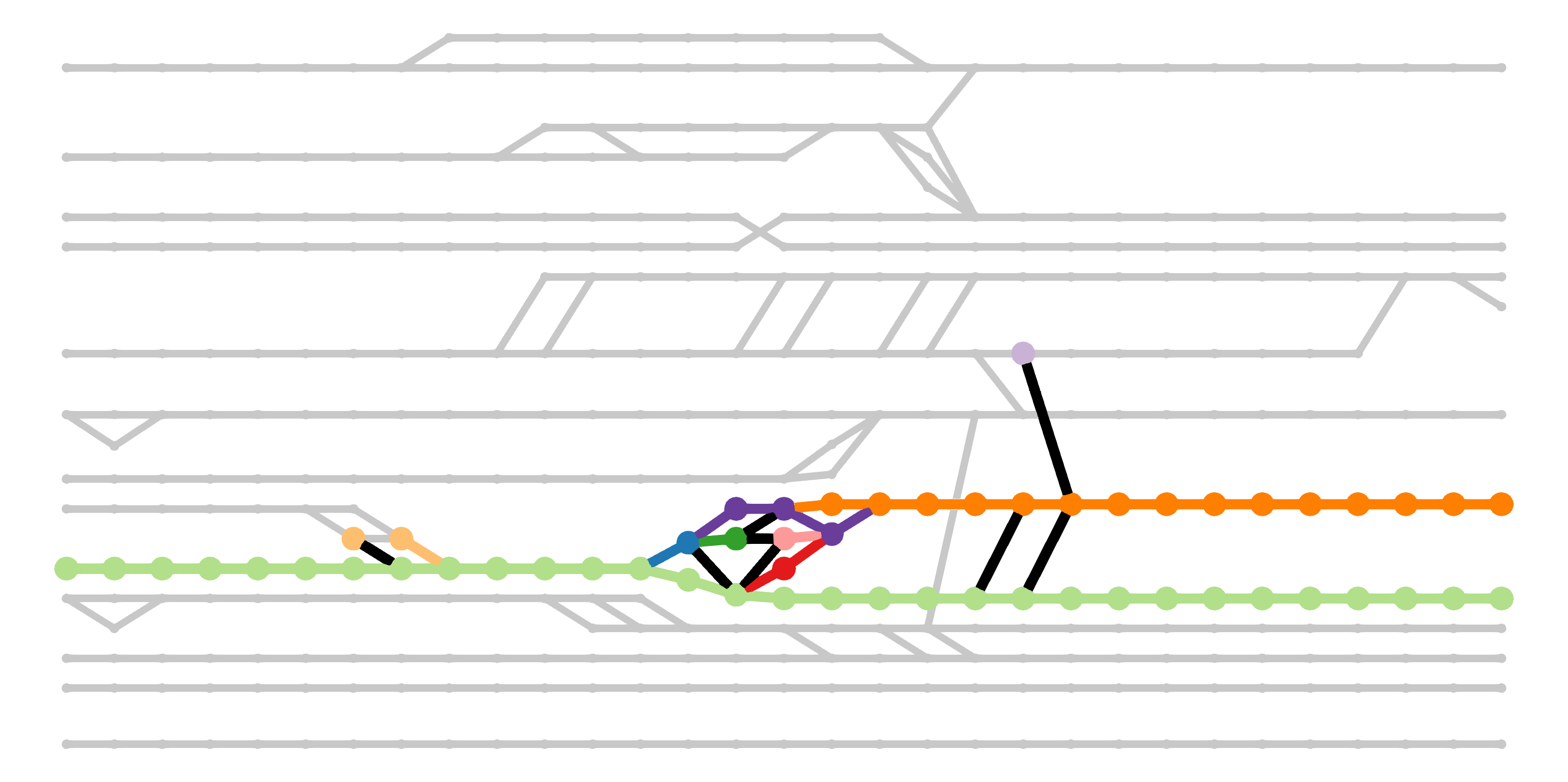}
    };
    
    \node(tg-xlabel)[below = 2.5 mm of graph] {\small \changed{(e)} \tiny Timestep ($t$) $\rightarrow$};

    \foreach \x [evaluate=\x as \xpos using (7.9-5.4) * (\x - 51)/6 + (5.4)] in {51, 52, ..., 57}{
        \node[scale=1] at (\xpos, 0) {\tiny \x};
        \draw[gray, line width=0.01mm] (\xpos, 0.2) -- (\xpos, 1.5);
        \draw[gray, line width=0.01mm] (\xpos,-1.5) -- (\xpos, -0.2);
    }

    \node[draw=black, line width=0.5pt, inner sep=1pt](closeup) [ above right =0.05 mm and 2.5mm of graph.2] {
        \includegraphics[width=3 cm, trim={40cm 11cm 46cm 34cm},clip]{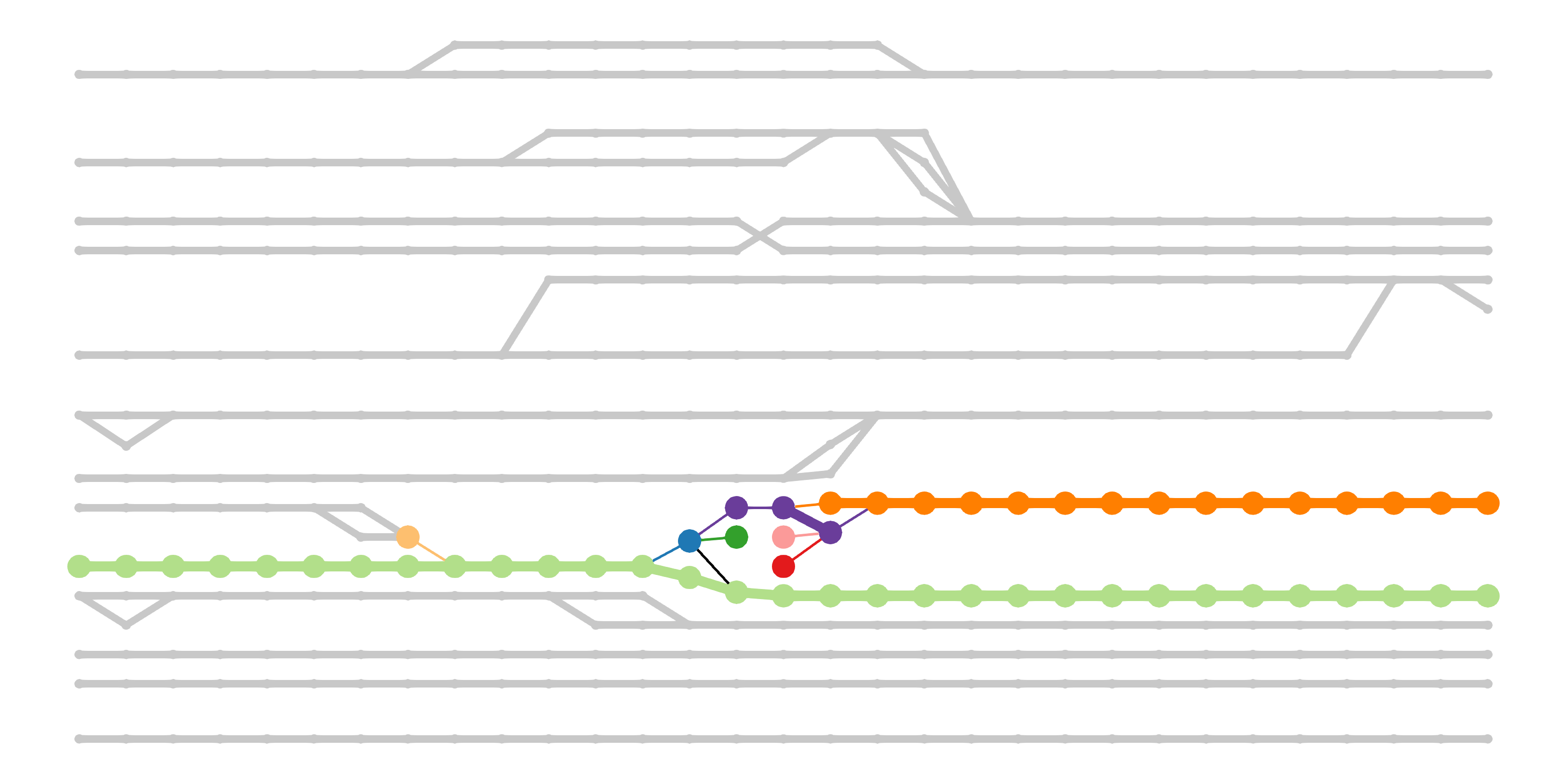}
    };

    \node[draw=black, line width=0.5pt, inner sep=1pt](manifold) [ below right =0.05 mm and 2.5mm of graph.358] {
        \includegraphics[width=3cm, trim={40cm 11cm 46cm 34cm},clip]{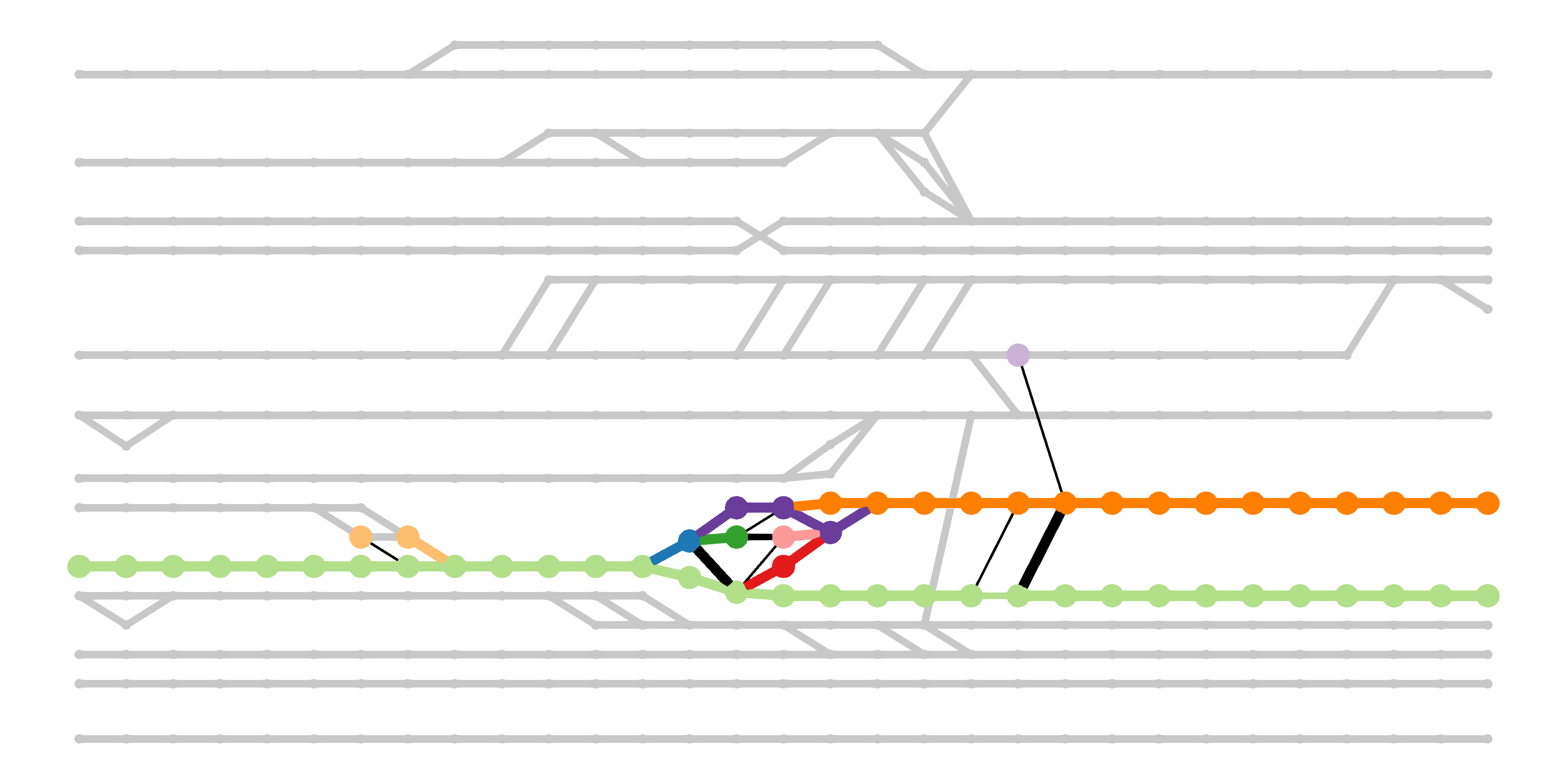}
    };

    \node[inner sep = 0.0pt, outer sep = 0.0pt, above right = 1mm and 0.5mm of closeup.0](closeup-lg1) {};
    \node[inner sep = 0.0pt, outer sep = 0.0pt, right = 4.5 mm of closeup-lg1](closeup-lg2) {};
    \node[inner sep = 0.0pt, outer sep = 0.0pt, below right = 1mm and 0.5mm of closeup.0](closeup-lg3) {};
    \node[inner sep = 0.0pt, outer sep = 0.0pt, right = 4.5 mm of closeup-lg3](closeup-lg4) {};

    \node[text=changed, scale=1, right = 0.001mm of closeup-lg2]{\tiny Unidirectional};
    \node[text=changed, scale=1, right = 0.001mm of closeup-lg4]{\tiny Bidirectional};
    
    \draw[black, line width=0.1mm] (closeup-lg1) -- (closeup-lg2);
    \draw[black, line width=1mm] (closeup-lg3) -- (closeup-lg4);

    \node[inner sep = 0.0pt, outer sep = 0.0pt, above right = 3mm and 0.5mm of manifold.0](manifold-lg1) {};
    \node[inner sep = 0.0pt, outer sep = 0.0pt, right = 4.5 mm of manifold-lg1](manifold-lg2) {};
    \node[inner sep = 0.0pt, outer sep = 0.0pt, above right = 1mm and 0.5mm of manifold.0](manifold-lg3) {};
    \node[inner sep = 0.0pt, outer sep = 0.0pt, right = 4.5 mm of manifold-lg3](manifold-lg4) {};
    
    \node[inner sep = 0.0pt, outer sep = 0.0pt, below right = 1mm and 0.5mm of manifold.0](manifold-lg5) {};
    \node[inner sep = 0.0pt, outer sep = 0.0pt, right = 4.5 mm of manifold-lg5](manifold-lg6) {};
    \node[inner sep = 0.0pt, outer sep = 0.0pt, below right = 3mm and 0.5mm of manifold.0](manifold-lg7) {};
    \node[inner sep = 0.0pt, outer sep = 0.0pt, right = 4.5 mm of manifold-lg7](manifold-lg8) {};

    \node[text=changed, scale=1, right = 0.001mm of manifold-lg2]{\tiny $< 25\%$ overlap};
    \node[text=changed, scale=1, right = 0.001mm of manifold-lg4]{\tiny $[25\%, 50\%]$ overlap};
    \node[text=changed, scale=1, right = 0.001mm of manifold-lg6]{\tiny $[50\%, 75\%]$ overlap};
    \node[text=changed, scale=1, right = 0.001mm of manifold-lg8]{\tiny $> 75\%$ overlap};

    \draw[black, line width=0.1mm] (manifold-lg3) -- (manifold-lg4);
    \draw[black, line width = 0.5mm] (manifold-lg5) -- (manifold-lg6);
    \draw[black, line width = 1mm] (manifold-lg7) -- (manifold-lg8);

    \node[below = 0.1mm of manifold] {\small \changed{(f)}};

    \node[draw=gray, inner sep=0pt, outer sep=0pt](t51) [above right = 2cm and 1mm of graph.180] {
        \includegraphics[width=3cm]{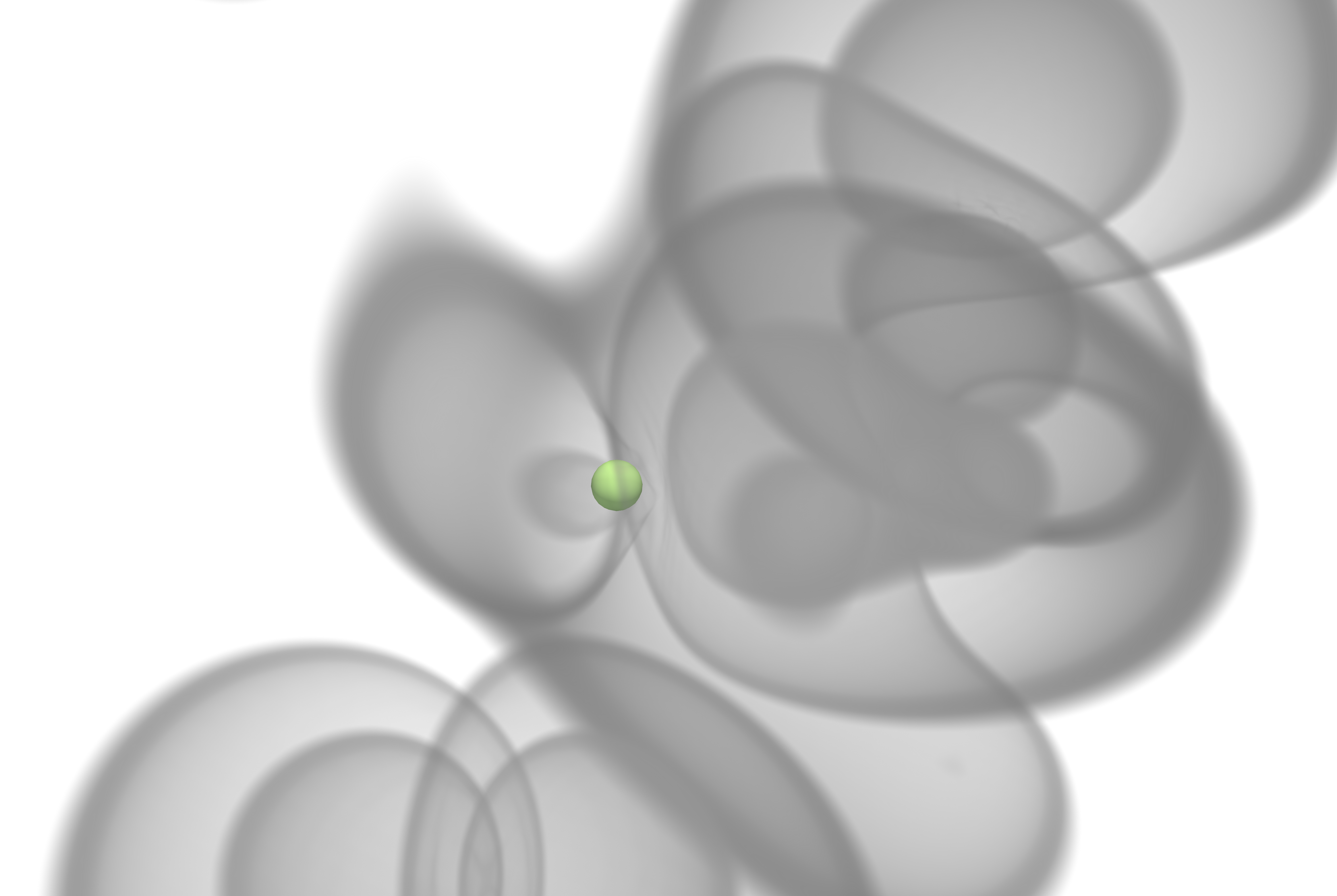}
    };
    \node[draw=gray, inner sep=0pt, outer sep=0pt](t53) [right = 0.5cm of t51] {
        \includegraphics[width=3cm]{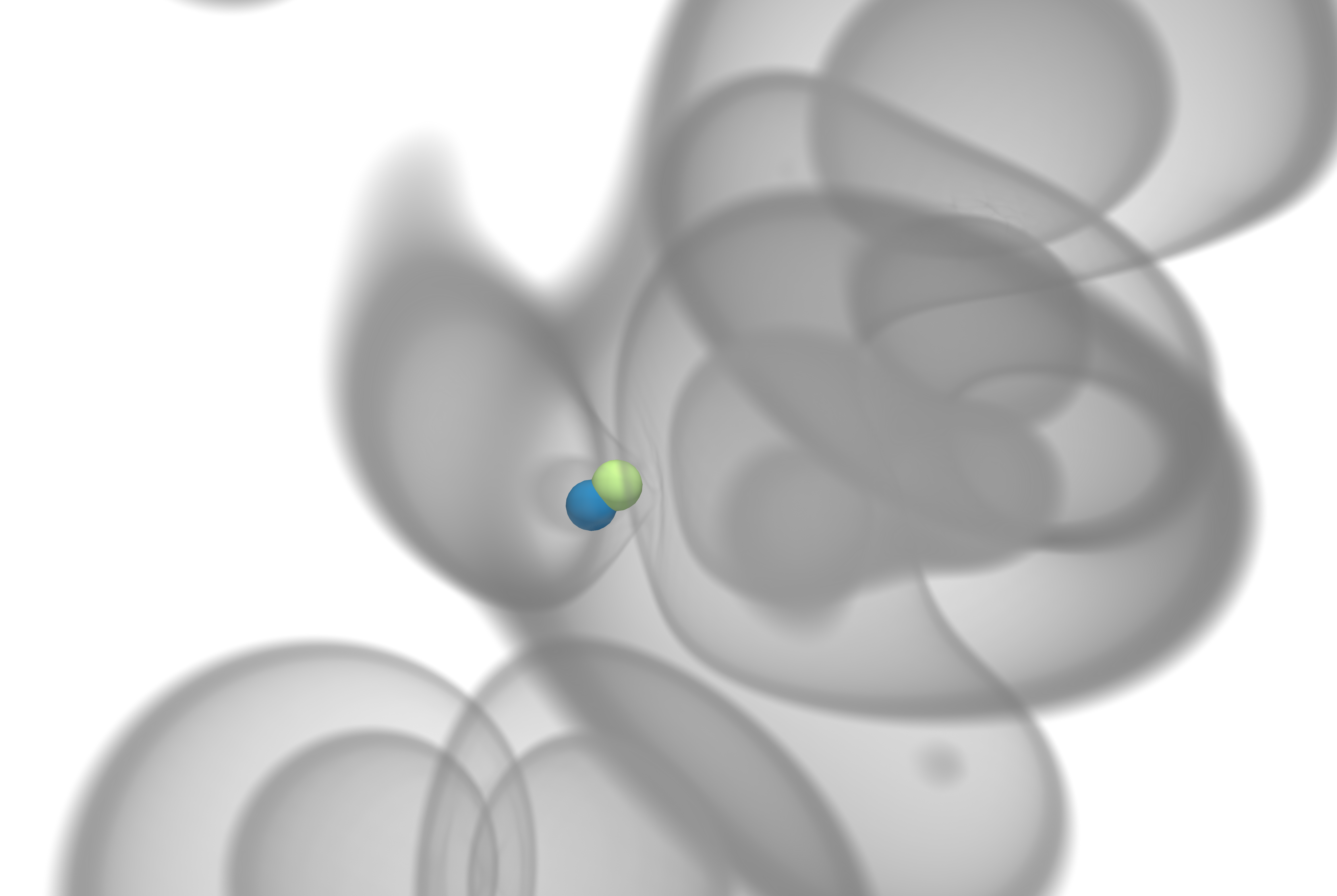}
    };
    \node[draw=gray, inner sep=0pt, outer sep=0pt](t55) [right = 0.5cm of t53] {
        \includegraphics[width=3cm]{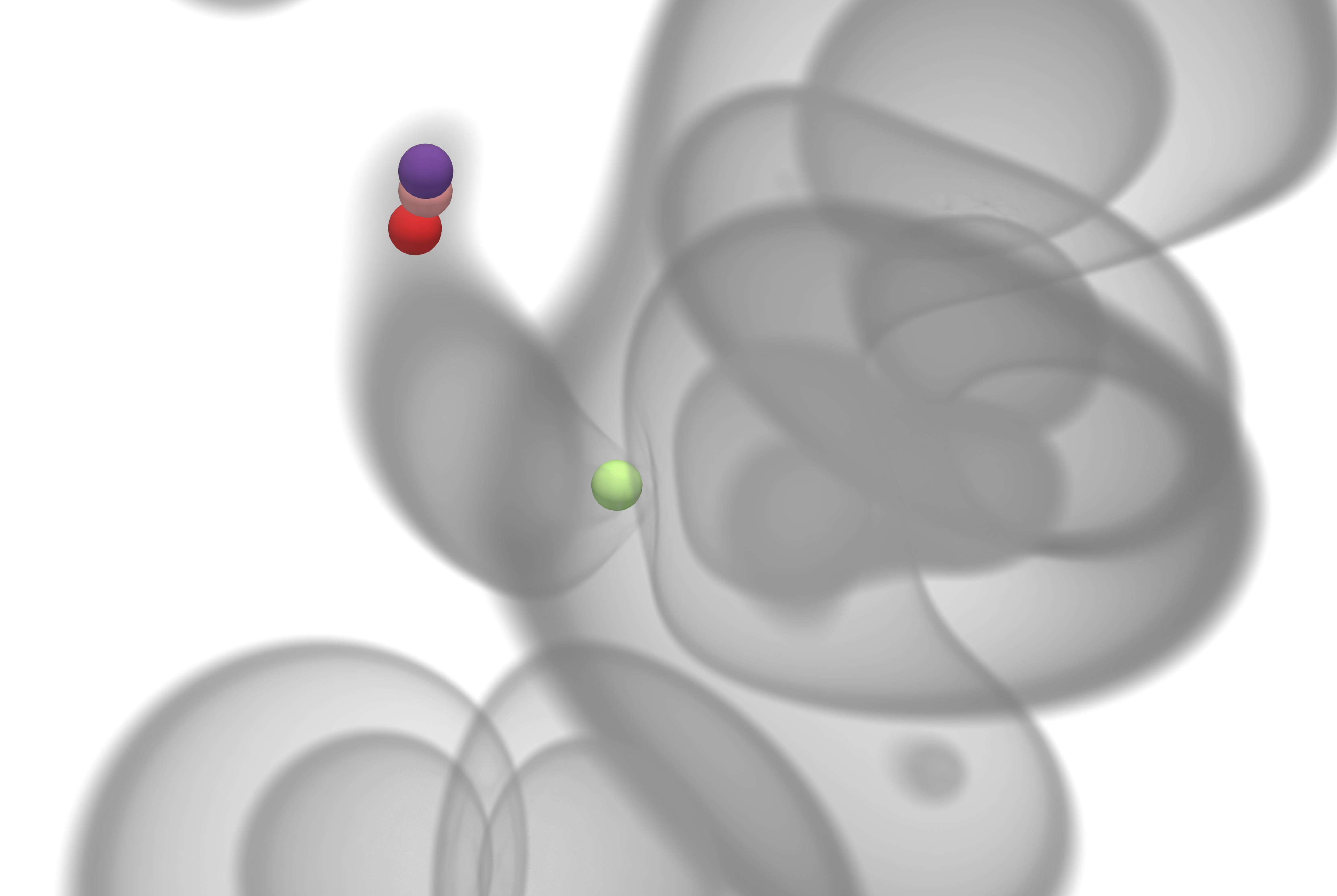}
    };
    \node[draw=gray, inner sep=0pt, outer sep=0pt](t65) [right = 0.5cm of t55] {
        \includegraphics[width=3cm]{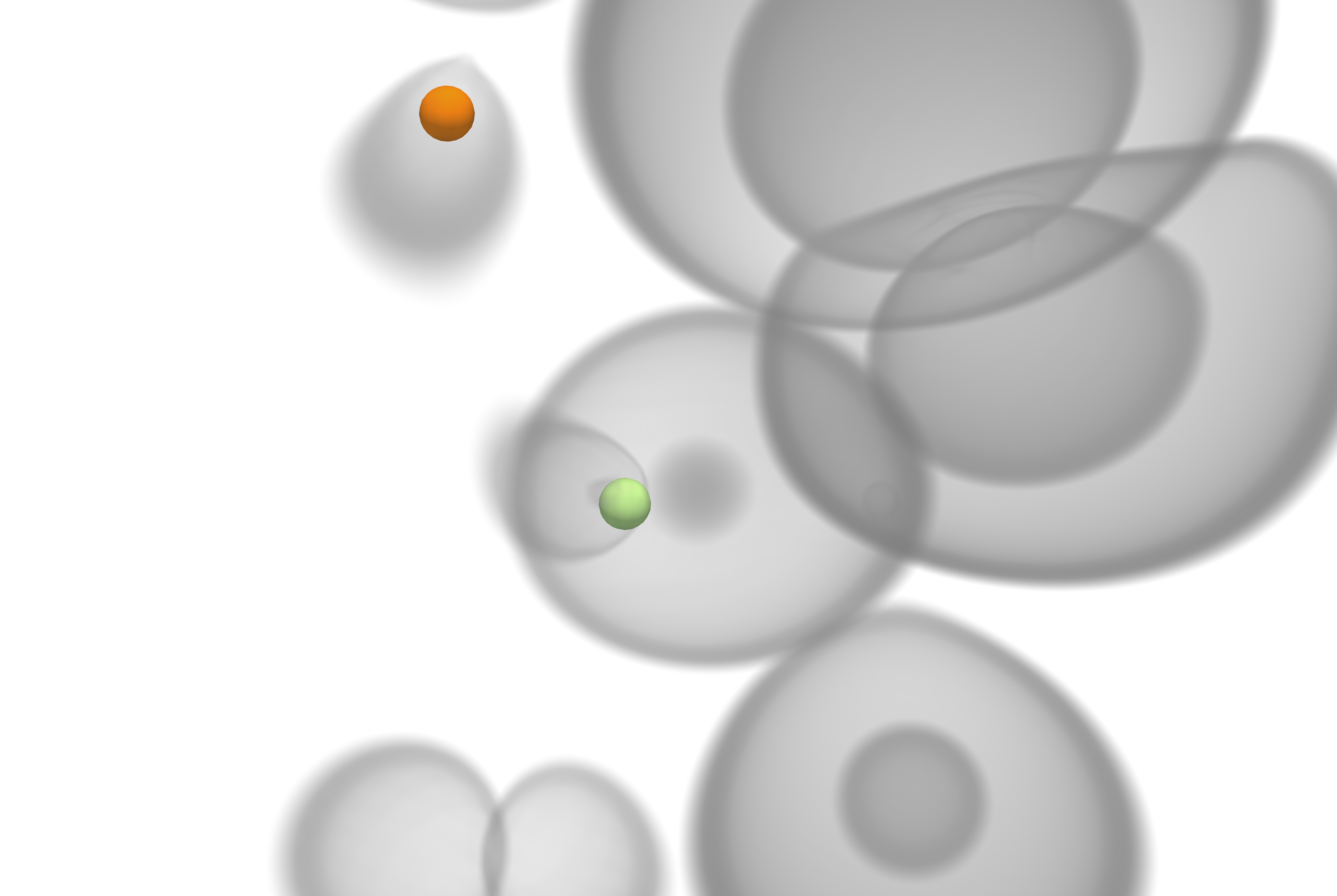}
    };

    \node[scale=1, below = 0.01mm of t51] {\small \changed{(a)} \tiny $t=$51};
    \node[scale=1, below = 0.01mm of t53] {\small \changed{(b)} \tiny $t=$53};
    \node[scale=1, below = 0.01mm of t55] {\small \changed{(c)} \tiny $t=$55};
    \node[scale=1, below = 0.01mm of t65] {\small \changed{(d)} \tiny$t=$65};  
\end{tikzpicture}%
    }
    \caption{The tracking graph computed for a molecular dynamics simulation showing the tracks of the maxima of the valence electronic density distribution using gradient-based tracking forward and backward in time. The tracking graph \changed{in (e)} gives an overview of how the electronic density over one part of the molecule changes between the time steps, while the top row \changed{(a-d)} shows volume renderings of the scalar field and the maxima at a selection of time steps. Each node and track of interest is colored and highlighted based on the index of the maximum. On the right \changed{in (f)}, we show two closeups of the tracking graph showing a split event in the features. \changed{On top, the line thickness is encoded by a one-to-one matching with two categories and on the bottom, the line thickness is instead encoded with four probability categories based on the proposed maximum descending manifold overlap between the maxima.}}
    \label{fig:electronic_density}
}

\abstract{
Feature tracking is a common task in visualization applications, where methods based on topological data analysis~(TDA) have successfully been applied in the past for feature definition as well as tracking.
In this work, we focus on tracking extrema of temporal scalar fields.
A family of TDA approaches address this task by establishing one-to-one correspondences between extrema based on discrete gradient vector fields.
More specifically, two extrema of subsequent time steps are matched if they fall into their respective ascending and descending manifolds.
However, due to this one-to-one assignment, these approaches are prone to fail where, e.g., extrema are located in regions with low gradient magnitude, or are located close to boundaries of the manifolds.
Therefore, we propose a probabilistic matching that captures a larger set of possible correspondences via neighborhood sampling, or by computing the overlap of the manifolds.
We illustrate the usefulness of the approach with two application cases. 
}


\begin{document}
\maketitle

\section{Introduction}
\label{sec:introduction}
%% ============================================================================
%% SECTION Introduction
%% ============================================================================
%
% Motivation
%
\begin{comment}
\begin{itemize}
    \item Gradient tracking is common and can work well (examples of applications)
    \item Previous methods do not encode how probable a match is given the gradient's behavior, \textcolor{red}{mostly visually evaluated? (at least how we did it in previous cyclone tracking paper)} Classical approach gives binary match.
    \item Semantically incorrect matches occur when an extrema is close to a ridge in the gradient field of the next/previous time step, as well as when a manifold in the field is large because of flat plateaus stemming from topological simplification. However, using discrete Morse theory requires us to simplify fields before processing them.
    \item Stability?
    \item Therefore, we explore techniques/methods yielding a probability for each match (short description of approaches).
\end{itemize}
\end{comment}

Feature tracking in scalar fields has become a standard application for topological data analysis~\cite{Yan2021}. Application examples span over many different domains, e.g., tracking of cyclones~\cite{Nilsson2022Cyclone}, vortices~\cite{Reininghaus2011,Saikia2017}, or burning structures in combustion simulations~\cite{bremer2009analyzing}. There is a wide variety of methods that consider different feature descriptors and different tracking concepts. All these methods have their strengths and weaknesses. In this work, we focus on features defined based on extrema in the scalar field combined with tracking based on discrete gradient vector fields. This means that two extrema of consecutive time steps are assigned to each other when they fall within their respective ascending and descending manifolds (basins). The result is a one-to-one extrema correspondence.
%in the forward and backward direction. 
The strength of this approach is that it is largely parameter-free (neglecting possible simplification thresholds) and is often successful even when features move widely between time steps compared to their ``size''. Size can be understood here as the height/depth of the extrema, but also as spatial extent when considering sub- or super-level-set representations. However, this approach also has its limitations and is prone to failure when extrema lie in regions of low gradient magnitude, or the neighborhood is dominated by strong extrema with large basins that attract most of the gradient lines. Other problems are related to the boundaries between basins, especially for data with low temporal resolution, a problem for many tracking methods. 

Since these extrema tracking methods are independent of the application and the semantic meaning of the scalar field, it is important to provide the domain scientist with the possibility to correct or reject the automatically generated assignments. Typically this is done by applying first some semantic filters and then visually checking the plausibility of the proposed tracks. To support this task we aim at generating more informative tracking results that provide a larger set of possible assignments in combination with an assignment weight. Since the one-to-one correspondence from gradient tracking cannot represent uncertainties, we propose to generalize gradient tracking by introducing probabilistic matching.
Specifically, we generate weights via neighborhood sampling, or by calculating the overlap of basins of consecutive time steps.
We demonstrate how this approach can be used to generate tracking graphs for hierarchical features defined as groups of extrema, e.g. according to topological simplification.
We apply our approach to two real-world applications, i.e., cyclone tracking, and tracking changes in electronic density fields.

\section{Related Work}
\label{sec:related}
%% ============================================================================
%% SECTION Related Work
%% ============================================================================
%

%\textcolor{red}{Maybe combine background and related work?}
\begin{comment}
\begin{itemize}
    \item common feature tracking methods for scalar fields (overlap, distance)
    \item gradient-based tracking, mention gradient tracking using manifolds by Engelke et al.~\cite{Engelke2020}, tracking the gradient path by Nilsson et al~\cite{Nilsson2022Cyclone}, Jacobi-Sets by Edelsbrunner et al. \cite{Edelsbrunner2002jacobi}. Also cite different ways to compute the ascending/descending manifolds \textcolor{red}{Robins paper}
\end{itemize}
\end{comment}

Below we briefly review some of the main concepts of topological features and their tracking. Since discussing all of the work in this area is beyond the scope of this overview, we focus on a few examples that illustrate the diversity of approaches. For a more detailed overview, we refer to the current report by Yan et al.~\cite{Yan2021}.

%topological feature descriptors
Various topological feature descriptors have been used in a variety of applications. The most common descriptors are based either on local extrema (maxima, minima, or both) or on sub-, super- or simply level sets. These two approaches are not entirely distinct, since all level sets contain at least one extremum. Unlike level sets, which require a scalar value to be specified, critical points are largely parameter-free if we neglect simplification during preprocessing. A concept, that falls somewhere in between, uses sets of critical points. An example is the so-called crown feature, which groups critical points of certain sub-branches of the merge tree. This corresponds to a set of extrema falling in a level set adding (or respectively subtracting) a local offset threshold from its dominant extremum. This feature definition has been used for cyclone definition~\cite{Engelke2020,Nilsson2022Cyclone} and also in the medical context of building a patient-specific brain atlas~\cite{Rasheed2022}. There are other more application-specific feature definitions using critical points. Examples of this are vortices in flow fields, which often use derived scalar fields for feature specification. Kasten et al.~\cite{Kasten2011,Kasten2016} consider minima in the acceleration field. Soler et al.~\cite{soler2018lifted} analyze ocean currents, eddies, and hurricanes based on critical point features. The use of level sets can also be found in many applications. Temperature level sets are used to analyze data from combustion simulations, often also in relation to other scalar fields, e.g. together with the fuel consumption~\cite{Weber2011}.
Topological features are often provided with additional attributes, such as the attached volume~\cite{Samtaney1994}.
%\todo{find better reference}.
A different type of feature definition can be obtained using extremal structures, which are sub-sets of the Morse-Smale complex~\cite{Narayanan2015,Homberg2014}.

% topological feature tracking
Tracking features over time or parameter space is one of the main tasks in many visualization applications. Depending on how the tracking dimension is handled, two different classes can be distinguished~\cite{Post2003}. In one class, features are defined directly in the spatio-temporal domain and the feature tracking task becomes a feature extraction task in a higher dimension. Another approach is to extract features per time step and then relate these features to each other in a second tracking step. To generate the final tracking graphs, often in an additional step, semantic filtering is performed, e.g., by excluding unrealistic large distances in the cyclone tracking work by Nilsson et al.~\cite{Nilsson2022Cyclone}, or by performing global match optimization~\cite{Schnorr2020}.

Since the generation of the feature correspondence for spatial features is the focus of our work, we will focus on this aspect in the following. Typically, this step applies some form of feature comparison resulting in a similarity or feature correspondence matrix~\cite{Nilsson2022Benchmark}. Various measures have been proposed for this comparison targeting different feature descriptors.
For features based on level sets, the most commonly used measures are spatial distance, spatial overlap, or attribute similarity.
Since level sets for different levels are nested, it is also possible to perform hierarchical tracking~\cite{Lukasczyk2019dynamic,Nilsson2020}, whose result can then be visualized with a nested tracking graph~\cite{Lukasczyk2017}.
For tracking extremal point-based features, global matching strategies considering some cost functions are often exploited. This includes the optimized matching between persistence diagrams based on the bottleneck or Wasserstein distances~\cite{Cohen-Steiner2010} or variants thereof~\cite{soler2018lifted}. Another common approach is to create a mapping between the merge trees. This includes several algorithms that calculate a tree edit distance~\cite{Sridharamurthy2020,Wetzels2022a}. While these methods are agnostic to the geometric embedding of the features, there are also attempts to combine topological distances with geometric properties~\cite{Yan2022b}.
In some examples, explicit temporal interpolation~\cite{Weinkauf2011} or optical flow~\cite{Valsangkar2019} was used to determine feature correspondence. A tracking method that also assumes linear interpolation between time steps utilizes Jacobi sets~\cite{Edelsbrunner2004}.
Point-to-point mapping can also be done using history-based tracking. For tracking in the visualization, this method was used as Combinatorial Feature Flow Fields as introduced by Reininghaus et al.~\cite{Reininghaus2011}. Since then, it has been modified and used in several applications~\cite{Kasten2012e,Engelke2020,Nilsson2022Cyclone}.
For a more detailed summary of topology-based tracking, see the review article by Yan et al.~\cite{Yan2021}.

% computation of ascending and descending manifolds: this should rather go into the background section

\section{Background}
\label{sec:background}
% ========================================================================================= %
% Background
% ========================================================================================= %
%\todoingrid{This section might need some attention \\
%What should be covered here? Critical points, which are our features. Gradient lines, AM/DM that we need for the tracking.\\
%In discrete Morse theory we son not consider a linear interpolation in the simplicial complex, this is an entirely discrete setting\\
%Strictly speaking, a discrete gradient line is not a list of vertices, but a list of simplices with alternating dimensions, differing by 1. \\
%In the description we might want to stick with the analytical case and then refer to some paper pointing at implementation, e.g. Vanessa Robins, and say is computed according to ..., and we represent it as an array of vertices?Maybe some more details about the paper [12] could be added.We should describe the original tracking approach here (forward/backward)}

Here we briefly recap key concepts from Morse theory~\cite{Matsumoto2002} which provides a mathematical foundation for the study of scalar field topology. 
For the sake of simplicity, we restrict our discussion to 2D scalar functions, however, the ideas generalize to higher dimensions. 
Given a smooth scalar function defined on a two-dimensional manifold $f:\mathbb{D}\to\mathbb{R}$, a point $c \in \mathbb{D}$ is called a \emph{critical point} if the gradient $\nabla f(c)$ at $c$ is zero. 
Furthermore, $c$ is a non-degenerate critical point if the matrix of second-order partial derivatives at $c$, also called Hessian, is invertible. 
The function $f$ is called \emph{Morse function} if it satisfies the following two conditions: all critical points are non-degenerate, and they have distinct scalar values. 
One of the key observations is that any smooth function can be infinitesimally perturbed to satisfy these two conditions, and hence key results from Morse theory are widely applicable to robust analysis of scalar fields encountered in a variety of scientific domains~\cite{bremer2009analyzing,shivashankar2015felix,Pandey2021}. 
The critical points of a Morse function can be classified based on the number of negative eigenvalues of the Hessian, also called the \emph{Morse index} of the critical point. 
For 2D scalar fields, the critical points are of three different types: minima, saddle, and maxima, with Morse index 0, 1, and 2, respectively. 

\begin{figure*}[t]
    \centering
    \subfloat[]{
        \includegraphics[width=0.24\linewidth]{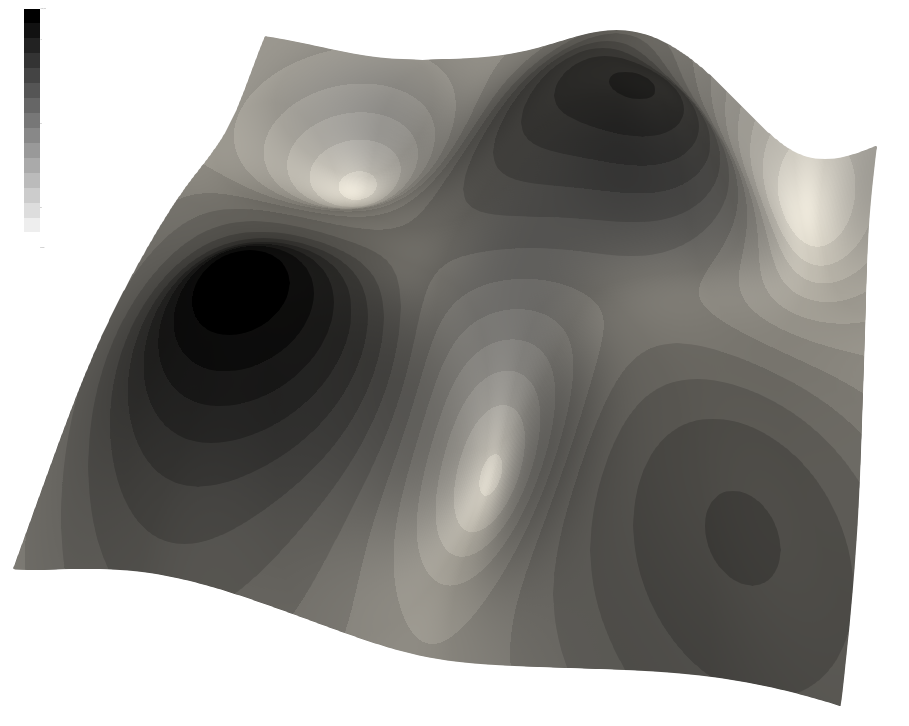}
        \label{subfig:input_sf}
    }
    \subfloat[]{
        \includegraphics[width=0.24\linewidth]{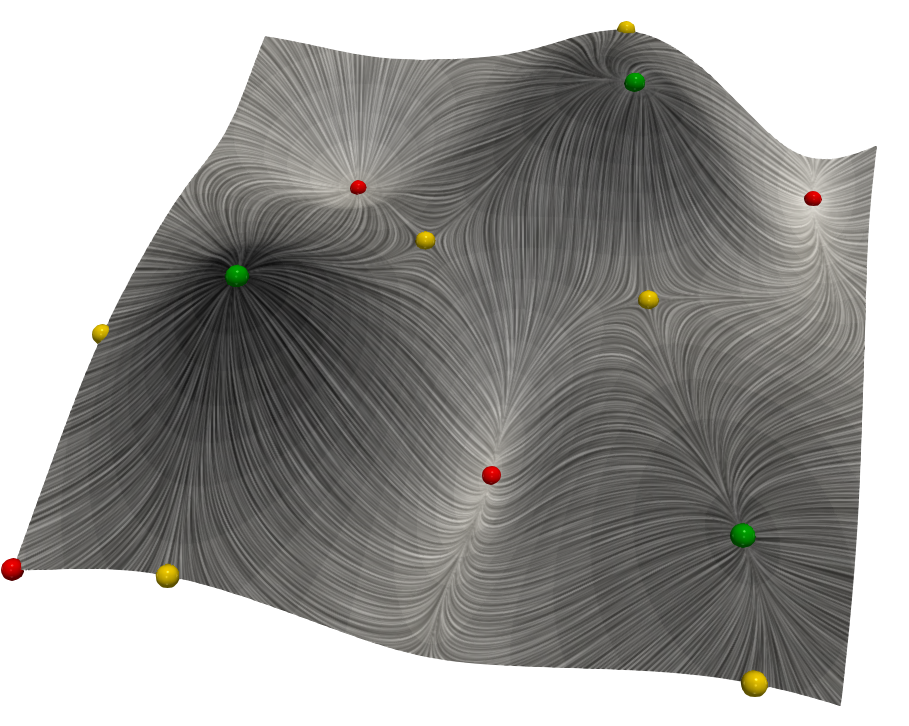}
        \label{subfig:lic}
    }
    \subfloat[]{
        \includegraphics[width=0.24\linewidth]{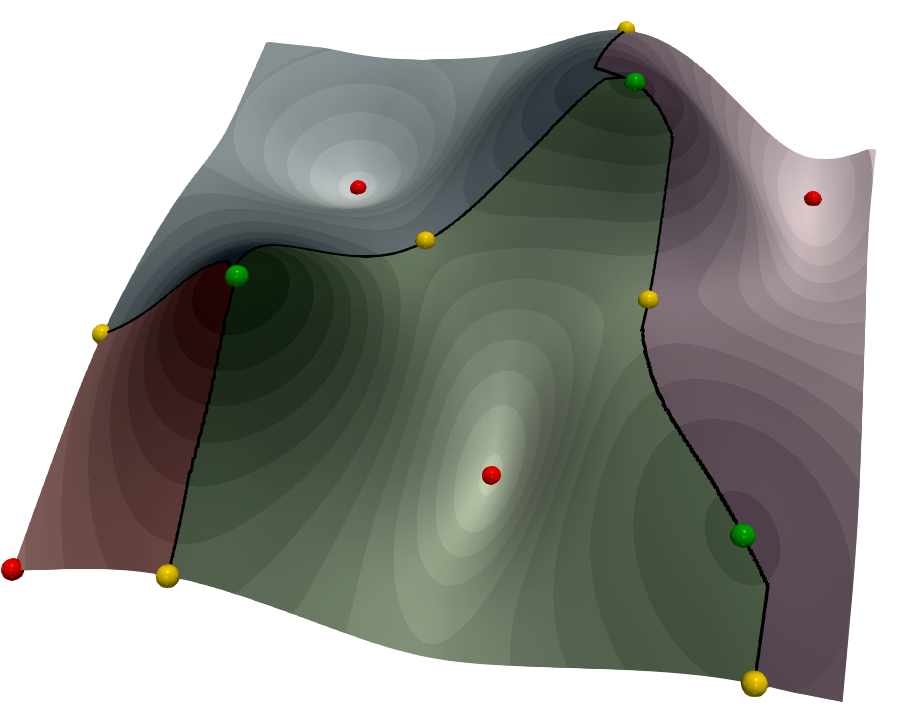}
        \label{subfig:asc_manifolds}
    }
    \subfloat[]{
        \includegraphics[width=0.24\linewidth]{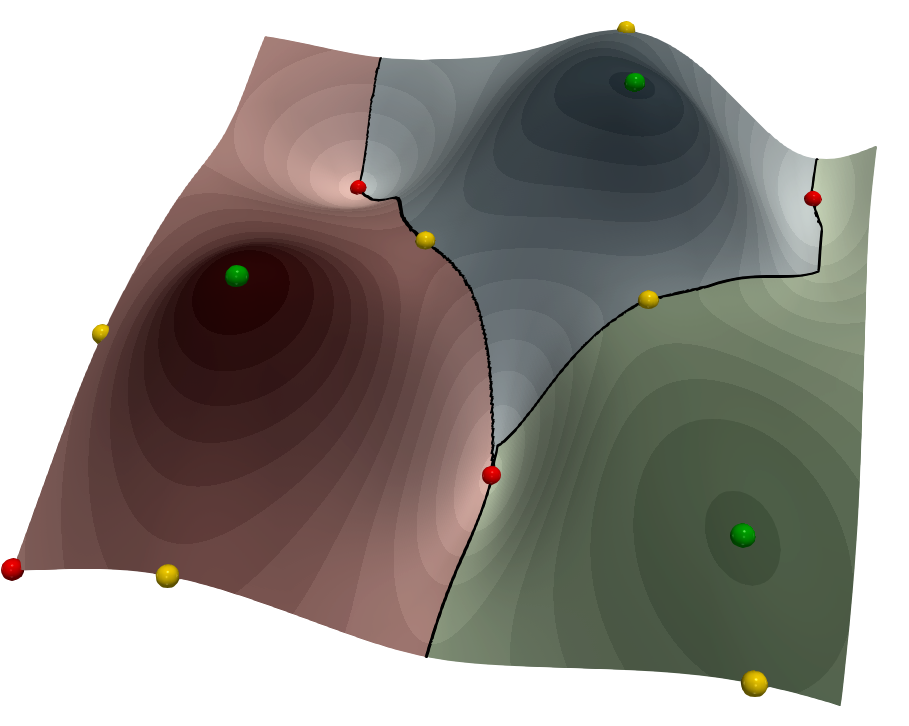}
        \label{subfig:desc_manifolds}
    }
    \caption{Illustration of the concept of descending and ascending manifolds. (a) shows the height field covered by scalar value, (b) shows a LIC image of its gradient vector field, and minima as red, saddles as yellow, and maxima as green spheres, (c) shows the ascending manifolds of the minima, and (d) the descending manifolds of the maxima.  }
    \label{fig:manifolds}
\end{figure*}

An \emph{integral line} is a curve in the domain $\mathbb{D}$ whose tangent at any point along the curve aligns with the gradient of $f$. 
In the limit, any integral line originates and ends at critical points. 
The set of all integral lines originating at a critical point $c$ along with the point $c$ is called the \emph{ascending manifold} $\aMan^c$ of $c$. 
Note that the dimension of the ascending manifold depends on its Morse index. 
An index $i$ critical point has an ascending manifold of dimension $2-i$ considering the case of 2D scalar fields.
The ascending manifolds of all critical points of $f$ partition the domain of $f$. 
This partition is called the \emph{Morse complex}~\cite{Edelsbrunner2003hierarchical,Shivashankar2012} of $f$ and consists of 2-cells corresponding to minima, 1-cells for saddles, and 0-cells for maxima. 
Symmetrically, one can study partitioning when we consider the endpoints of the integral lines rather than their origin. 
In this case, the corresponding manifold for a critical point $c$ is called \emph{descending} manifold $\dMan^c$ and its dimension is the same as the index of $c$.

\autoref{fig:manifolds} illustrates these concepts on a synthetic 2D scalar field. \autoref{subfig:input_sf} shows the 2D manifold $\mathbb{D}$ with the height function as the associated scalar field $f$. In \changed{\autoref{subfig:lic}}, the gradient field $\nabla f$ is shown using standard vector field visualization method called Line Integral Convolution~(LIC). The critical points are also shown as colored spheres, with red, yellow, and green spheres denoting minima, saddles, and maxima, respectively. This image also provides a very good perception of the integral lines in $\mathbb{D}$, and how they emerge out of minima and converge at maxima. \autoref{subfig:asc_manifolds} shows the segmentation or partitioning of the domain $\mathbb{D}$ induced by the ascending manifolds of the critical points. The four minima partition the domain into four regions separated by ridges shown as thick black curves connecting the saddles and maxima. An ascending manifold of a minima can also be thought of as a valley or the watershed basin associated with the minimum. Lastly, \autoref{subfig:desc_manifolds} shows the symmetric segmentation of the domain based on descending manifolds of the three maxima.

So far we have discussed Morse theory and the associated concepts for smooth functions defined on manifolds. 
However, in practice, we rarely find such functions. 
Instead, we are provided discrete samples of such functions on a point cloud sample of $\mathbb{D}$. 
On this point cloud, a combinatorial structure called a \emph{simplicial complex} $\domain$ is often defined as an approximation of $\mathbb{D}$. 
Therefore, the input is a piecewise linear scalar field $f: \domain \rightarrow \R$ with real values given at the vertices $\vertices$, and values inside higher dimensional simplices are linearly interpolated. 
%The input of our approach is a piecewise linear scalar field $f: \domain \rightarrow \R$ defined on a $d\leq3$ dimensional simplicial complex $\domain$, where real values are given at the vertices, and values inside higher dimensional simplices are linearly interpolated.
%Following discrete Morse theory~\cite{Forman2001}, we consider the gradient $\nabla f$ to be based on the discrete gradient vector field of $f$.
In this piecewise linear setting, an integral line originating at any vertex $v_0\in\vertices$ can be approximated by the line of steepest ascent defined as a sequence of vertices $(v_0,v_1,\dots,v_n)$, where $v_{i+1}$ is the largest neighbor of $v_{i}$ for all $i\in[0,n)$, and $v_n$ is a maximum of $f$ i.e. a vertex that has no larger neighbors.
To disambiguate vertices with the same scalar value, we apply \emph{Simulation of Simplicity}~\cite{Edelsbrunner1990simulation}; which enforces that there is always a unique largest neighbor by, e.g., breaking ties based on vertex indices.
%
%Then we store the descending manifold as a map $\dMan:\vertices\rightarrow\vertices$ that assigns to each vertex the endpoint (maximum) of its corresponding integral line.
The descending manifolds $\dMan^{m_i}$ of the set of maxima $m_i\in\mathcal{M}$, therefore, segments the domain into regions with uniform gradient flow behavior.
Symmetrically, we define the ascending manifold $\aMan$ based on tracing the integral lines in the direction of steepest descent, which therefore terminate at minima.
% where all vertices $v$ have integral lines which originate from the same minima of $f$ and have the same maxima as destination.
% In the smooth setting, an \emph{integral line} $I_f$ is a curve whose tangent at every point $v\in I_f$ is parallel to the gradient $\nabla f(v)$ of $f$ at $v$.
% Also note that $f$ is made injective by simulation of simplicity \cite{Edelsbrunner1990simulation}, where each vertex $v$ is perturbed to have a different value $f(v)$ so the entire field can be ordered, \textcolor{red}{which is a requirement for using methods from discrete Morse theory}.
% Extrema are types of critical points, the minima and maxima of $f$, that are locally classified according to the neighboring vertices. A minima $m^-$ is a vertex $v$ with lower $f(v)$ than all its neighbours and vice versa for maxima $m^+$. Saddle points are also critical points, however they are not trackable with our appraoch as we focus solely on extrema. Finally, the field also consists of regular vertices $v$ which are vertices within the contour components. Contour components are level sets, sublevel sets or superlevel sets. Consider an iso-value $i$ of the scalar field $f$. A level set is defined as $L(i) = \{v \in \domain | f(v) = i\}$, a sublevel set as $L(i) = \{v \in \domain | f(v) < i\}$ and a superlevel set as $L(i) = \{v \in \domain | f(v) > i\}$.

Gradient-based tracking approaches utilize the ascending (descending) manifolds for establishing the mapping between minima (maxima) of consecutive time steps. So, a minimum $m_i$ of the scalar field $f_t$ corresponding to time step $t$ is assigned to a minimum $m_j$ of $f_{t+1}$ if the position of $m_i$ is contained within $\aMan^{m_j}$. In this way a forward mapping is established between the set of minima in time step $t$ to those in $t+1$. Note that there is no flexibility in this assignment, a minimum is assigned to a unique minimum in the next time step. Also note that the forward mapping can only capture continuation and merge events. To address some of these challenges, a backward map is also constructed that assigns a unique minimum in $f_{t-1}$ to each minimum in $f_t$. Now, split events can also be captured and we have some notion of strength of connection between minima in consecutive steps, with an indication of ``strong'' connection between $m_i$ of $m_j$ if the they are mapped to each other via the forward and backward maps, and ``weak'' connection if only one of the forward or backward map establishes a connection between the pair of minima. These gradient-based tracking approaches either explicitly compute the integral lines of point features~\cite{Nilsson2022Cyclone}, or retrieve that information from precomputed  ascending/descending manifolds~\cite{Engelke2020}.
Our approach follows the latter, where the manifolds can efficiently be computed via path compression~\cite{Maack2023MSSegmentations}.

\section{Method overview}
\label{sec:method}

A basic assumption in our work is that the features of interest can be associated with either maxima or minima in a scalar field that can be followed over time. 
In the rest of the paper, we use extrema as a general term to refer to one type of extrema, i.e., either maxima or minima. 
Conceptually, we assume that we can model the tracking in two parts. 
The first part is the basic tracking of extrema, which should be generic with a minimal number of adjustable parameters and user interactions, resulting in a raw tracking graph.
In the second part, this raw tracking graph is adapted to a specific application. This step allows the domain scientist to set specific preferences. This means providing a wide range of options to compose the final tracks, including filtering and correcting the tracking results as well as adjusting the level of detail of the features.
In this work, we mainly focus on the first part, but we need to keep the second part in mind when making decisions about metadata that needs to be stored in our raw tracking results.
In the following, we first give an overview of the pipeline for the first part. 
The summary of all notations can be found in the \autoref{sec:notations}.

%\subsection{Pipeline overview}

\paragraph{Input} A time series of piecewise linear scalar fields $f_{t}$ for time steps $ t=1\dots N$, typically defined over a 2- or 3-dimensional domain $\domain$. Additionally, the type of extrema to be tracked is also specified.
\paragraph{Extrema extraction} Computation of extrema $m_i\in\mathcal{M}$
%$m_j\in \mathcal{M}_{t}$ 
for each time step representing our raw features. To keep the basic tracking as parameter-free as possible, we only apply minimal simplification at this point to remove topological noise, \changed{but for specific applications the parameter can be tuned based on inputs from domain scientists}. If hierarchical tracking should be supported, a clustering hierarchy of extrema is also computed for each time step. This clustering hierarchy can, for example, be computed based on a branch decomposition of the merge tree, crown features, or persistence simplification of extremum graphs.
\paragraph{Correspondence generation} The sets of extrema of consecutive time steps are related to each other using different variants of probabilistic gradient tracking\changed{, where the computation of the ascending/descending manifold is the most expensive part of the algorithm}. The result of this step is a set of two correspondence matrices, one for backward correspondence $\Cbackward(t)$ and one for forward correspondence $\Cforward(t)$. The matrix rows represent the extrema from the current time step, and the matrix columns represent the extrema of the previous timestep in $\Cbackward(t)$ or the next time step in $\Cforward(t)$. 
The entries in the matrix store the probability of a connection between the extrema.
These correspondence matrices are computed based on overlap matrices $\Overlap^\pm(t)$.
\paragraph{Output} Sets of extrema $\mathcal{M}_{t}$, the clustering hierarchy if required, the forward and backward overlap matrices $\Overlap^\pm(t)$ and the final correspondence matrices  $\Cmatrix^\pm(t)$ for all time steps $t$.

%\todoingrid{We should probably start with a summary of the tracking steps and introduce the different matrices involved and how they are used. What is the input what is the output, and what do we expect from our "method"?}

\section{Probabilistic gradient tracking}
\label{sec:tracking}
%\section{Probablistic gradient tracking}
%\label{sec:tracking}

\begin{figure}[t]
    \centering
        \adjustbox{width=\linewidth, valign=b}{
          \input{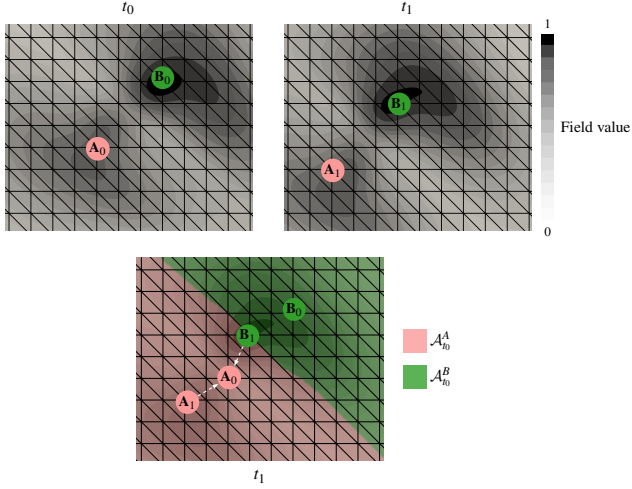}
        }
        \vspace{-1em}
    \caption{Case ridge: In the top row, a time-dependent scalar field is shown at $t_0$ and $t_1$ respectively, where there are two minima A and B in $t_0$ and $t_1$. To track $A_1$ and $B_1$ in $t_0$ with gradient-based tracking, we can calculate the ascending manifolds $\aMan_{t_0}$ of $t_0$, see the lower part of the figure. Both $A_1$ and $B_1$ fall within the ascending manifold $\aMan^A_{t_0}$, as $B_1$ has fallen on the other side of the ridge between its actual corresponding minima $B_0$.}
    \vspace{-1.4em}
    \label{fig:ridge}
\end{figure}

\begin{figure*}[t]
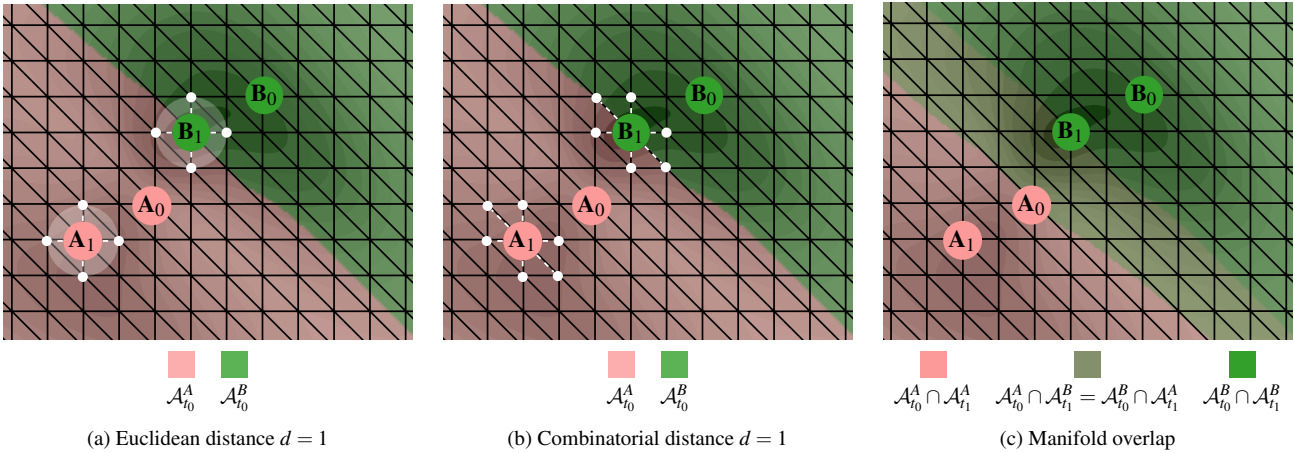

    \centering
    \subfloat[Euclidean distance $d=1$]{
        \adjustbox{width=0.32\linewidth, valign=b}{
          \input{Figures/illustrations/illustration_casesv2/method-euclidean}
        }
     \label{subfig:ridge-dist}
    }
    \subfloat[Combinatorial distance $d=1$]{
        \adjustbox{width=0.32\linewidth, valign=b}{
          \input{Figures/illustrations/illustration_casesv2/method-combi}
        }
     \label{subfig:ridge-combi}
    }
    \subfloat[Manifold overlap]{
        \adjustbox{width=0.32\linewidth, valign=b}{
          \input{Figures/illustrations/illustration_casesv2/method-manifold}
        }
     \label{subfig:ridge-manifold}
    }
    \caption{In this figure we illustrate how our methods access the gradient to encode probability in the matches, based on the example in \autoref{fig:ridge}. 
    In a), we sample the neighborhood (white points) around $A_1$ and $B_1$ using Euclidean distance, to determine the best match. For $A_1$ this yields a 100\% likely match to $A_0$. For $B_1$ an equal number of sample vertices fall into $\aMan_{t_0}$ and $\bMan_{t_0}$, however since $B_1$ falls in $\aMan_{t_0}$, \changed{the match to $A_0$} is more probable. In b), we use the combinatorial sampling, for the distance $d=1$. In this case, $A_1$ is fully matched with $A_0$, while $B_1$ is still matched with both $A_0$ and $B_0$, with a higher probability for $A_0$.
    Finally, in c)  we consider the overlapping manifolds of the fields at $t_0$ and $t_1$. In this figure, the ascending manifolds of both time steps are overlayed, the gray area shows the overlap between $\aMan_{t_0}$ and $\bMan_{t_1}$, while the pink and green areas are the overlap between $\aMan_{t_0}$ and $\aMan_{t_1}$ and $\bMan_{t_0}$ and $\bMan_{t_1}$, respectively. With this approach, $B_1$ has a higher probability of matching with $B_0$ than $A_0$, as the overlap is larger.}
    \label{fig:ridge-methods}
\end{figure*}

In the following section, we define two novel strategies that encode the probability of an extremum in the scalar field $f_t$ being matched with extrema in the scalar fields $f_{t-1}$ and $f_{t+1}$, respectively. Both of our approaches are based on the flow of the gradient in relation to the extremum. 
We use the ascending respective descending manifolds for minima resp. maxima tracking.
By using more information from these manifolds than the current location of the extrema, we get a larger set of possible correspondences.

We use a running example to illustrate our different approaches shown in \autoref{fig:ridge}.
%To motivate probabilistic tracking, we give an example of a case where simple gradient-based tracking approaches can fail. 
%In the top row of \autoref{fig:ridge}, 
The example consists of two scalar fields for $t=t_0$ and $t=t_1$ respectively with two minima  $A$ and $B$ in each time step. To track $A$ and $B$ backward in time, e.g. from $t_1$ to $t_0$, we use the ascending manifolds of the minima. In the bottom image \autoref{fig:ridge}, the ascending manifolds $\aMan^A_{t_0}$ and $\aMan^B_{t_0}$ are overlayed on top of the scalar field at $t_1$. Here, $B_1$ and $A_1$ both fall within $\aMan^A_{t_0}$, and are given a one-to-one correspondence, despite $B_1$ lying close to \changed{the boundary between the two ascending manifolds $\aMan^A_{t_0}$ and $\aMan^B_{t_0}$}. In the following section, we illustrate how our proposed strategies can generate correspondences where $B_1$ is also matched with $B_0$.

\subsection{Sampling local neighborhood}
Our first approach samples vertices in the neighborhood of the local extrema to determine the correspondence to the extrema in the next or previous time step.
To define the sampling neighborhood, we first define a distance threshold $d$ and a distance function $D(v_i,v_j)$ that assigns a distance to each pair of vertices in the domain. Then the local neighborhood vertex set $\domain_\vertices$ of the extremum $m$ is defined as
\begin{equation}
    \domain_\vertices(m) = \{v \in \vertices | D(m, v) \leq d\}
    \label{eq:sampling-neighbourhood}
\end{equation}
where $\vertices\subset \domain$ is the set of vertices in the domain $\domain$.
%, $m$ is the extremum in question.
The cardinality of the set $\domain_\vertices(m)$ is denoted $|\domain_\vertices(m)|$.
We explore two possible distance functions: the Euclidean distance, or the combinatorial distance in the triangulation. 
Consider two vertices $v_i, v_j \in \vertices$, the Euclidean distance is simply $|| v_i - v_j||_2$, whereas
%, and the distance threshold $d$ can be any real-valued number. 
%
the combinatorial distance $D(v_i, v_j)$ is defined as the length of the shortest path $(v_i, v_{i+1}, ... , v_j)$ in the triangulation. 
%that minimizes the number of edges $e(v_i, v_{i+1})$ in $P$. 
%In this case, the distance threshold is given as a number in $\N$. 
To compute $\domain_\vertices(m)$, we utilize a breadth-first search approach, iteratively adding neighboring vertices $v$ to $\domain_\vertices(m)$ as long as $D(m, v)$ is smaller than or equal to $d$.
%we sample the local neighborhood of $m$ by iteratively querying the star $\Star_\vertices(\domain_\vertices(m))$. 
 %For each vertex $v_i \in \domain_\vertices(m)$ get the star $\Star_\vertices(v_i)$. 
%For each vertex $v_j \in \Star_\vertices(v_i)$ add $v_j$ to $\domain_\vertices(m)$ if $D(m , v_j) \leq d$. 

The two sampling strategies are illustrated in \autoref{fig:ridge-methods} using the same example as in \autoref{fig:ridge}. The results, in this case, are very similar. $B_1$ corresponds to both $A_0$ and $B_0$, and $A_1$ still only corresponds to $A_0$.
%For the combinatorial distance, \autoref{subfig:ridge-combi}  $B_1$ is matched with both $A_0$ and $B_0$, and $A_1$ with $A_0$.
%
During this process, we count how many of the samples from $\domain_\vertices(m_i)$ in time step $t$ fall into the manifolds of the extrema $m_j$ in $f_{t-1}$ respective $f_{t+1}$ stored as \emph{overlap values}.
For each pair of extrema, $m_i\in \mathcal{M}_{t}$ and $m_j\in\mathcal{M}_{t\pm 1}$  we define forward/backward overlap values as
\begin{equation}
\Overlap^\pm_{i,j}(t) =  |\domain_\vertices(m_i)\cap \dMan_{t\pm1}^{m_j}| 
\end{equation}
These values are stored in \emph{overlap matrices} $\Overlap^\pm(t)$ of size $|\mathcal{M}_{t}|\times|\mathcal{M}_{t\pm 1}|$.
The \emph{correspondence matrices} $\Cmatrix^\pm(t)$ are derived from the overlap matrices by normalizing with respect to the number of vertices in the set $|\domain_\vertices(m_i)|$.
\begin{equation}
\Cmatrix^\pm_{i,j}(t) =  \frac{\Overlap^\pm_{i,j}(t)}{|\domain_\vertices(m_i)|}
\end{equation}
%a correspondence map for both temporal directions, $\CforwardMap: \N \rightarrow \N$ and $\CbackwardMap: \N \rightarrow \N$. The keys of the maps are the indices of the manifolds we match with, represented by the minimum/maximum vertex index, while the values are the number of $v \in \domain_\vertices(m)$ that fall into that manifold. 
%Finally, the probability of each match is derived normalizing these values by the number of vertices in the set $\domain_\vertices(m)$. 
In conclusion, the probability of a match is the proportion of vertices in the local neighborhood of the current extremum $m_i$ that falls within the manifold of an extremum in the previous time step $m_{t-1}$ or following time step $m_{t+1}$, which is in accordance with the definition of conditional probability.

%\todoingrid{What is a good value for d? Could we suggest a rule of thumb?}
%\todoemma{I haven't tried to study what a good value for $d$ would be, so I don't think it's possible to suggest something }

\subsection{Manifold overlap}
In the second approach, we take the entire manifold into consideration. In contrast to the sampling approach, manifold overlap is parameter-free. Recall that for the descending manifold $\dMan$, all vertices $v \in \dMan$ have the same maximum as the endpoint of their gradient line, while for the ascending manifold $\aMan$, all vertices $v \in \aMan$ have the same minimum as the start point of their integral line. Therefore, each manifold in a scalar field can be represented by the extremum $m$ where the gradient lines end/start. In the following, we will restrict our description to maxima and descending manifolds. The situation is analogous for minima and ascending manifolds.

In our strategy, we first compute the size of the descending manifolds $|\dMan_t^{m}|$ for all maxima $m$ by counting the number of vertices, where the corresponding maximum to the manifolds are provided to the tracking. 
Then we calculate the intersection of all manifolds in consecutive time steps.
For example, for a maximum $m_i$ at time $t$ and a maximum $m_j$ at time $t+1$ their overlap is the intersection of their descending manifolds $\dMan_t^{m_i} \cap \dMan_{t+1}^{m_j}=\{v\in\domain| v \in \dMan_t^{m_i} \wedge v \in \dMan_{t+1}^{m_j}\}$. 
We store the size of these intersections 
%$|\dMan_t^{m_i} \cap \dMan_{t+1}^{m_j}|$ 
in \emph{overlap matrices} $\Overlap^\pm(t)$ for all consecutive time steps with the entries $(i,j)$ given by
\begin{equation}
\Overlap^+_{i,j}(t) = \Overlap^-_{i,j}(t+1) = |\dMan_t^{m_i} \cap \dMan_{t+1}^{m_j}|.
\end{equation}
These matrices contain all the information needed for the forward and backward correspondence matrices $\Cmatrix^\pm(t)$.
Therefore, the probability is encoded by normalizing the overlap with the size of the current descending manifold, which again is in line with the conditional probability.
Then correspondence matrix entries for $m_i$ at time $t$ are then defined as 
\begin{equation}
\Cmatrix^\pm_{i,j}(t) = \frac{\Overlap^\pm(t)_{i,j}}{|\dMan_t^{m_i}|} = \frac{|\dMan_t^{m_i} \cap \dMan_{t\pm1}^{m_j}|}{|\dMan_t^{m_i}|}.
\end{equation}
%and the forward manifold overlap probability as
%\begin{equation}
%\Cbackward_{i,j}(t) = \frac{|\dMan_t^{m_i} \cap \dMan_{t-1}^{m_j}|}{|\dMan_t^{m}|}.
%\end{equation}

Returning to our example \autoref{fig:ridge}, we show the result of using this strategy \changed{in \autoref{subfig:ridge-manifold}}.
We have now overlayed the scalar field from $t_1$ with all four ascending manifolds $\aMan^A_{t_0}$, $\aMan^B_{t_0}$, $\aMan^A_{t_1}$ and $\aMan^B_{t_1}$. The strong pink and green regions respectively shows the overlap $\aMan^A_{t_0} \cap \aMan^A_{t_1}$ and $\aMan^B_{t_0} \cap \aMan^B_{t_1}$, while the green-gray region is the overlap between $\aMan^A_{t_0} \cap \aMan^B_{t_1} = \aMan^B_{t_0} \cap \aMan^A_{t_1}$. In this case, $A_1$ is matched to both $A_0$ and $B_0$, just like $B_1$ was in the previous cases. However, for the current example, $A_1$ and $B_1$ will have more probable correspondences to $A_0$ and $B_0$ respectively.

\begin{figure*}[t]
    \centering
    \adjustbox{width=\linewidth}{
          \input{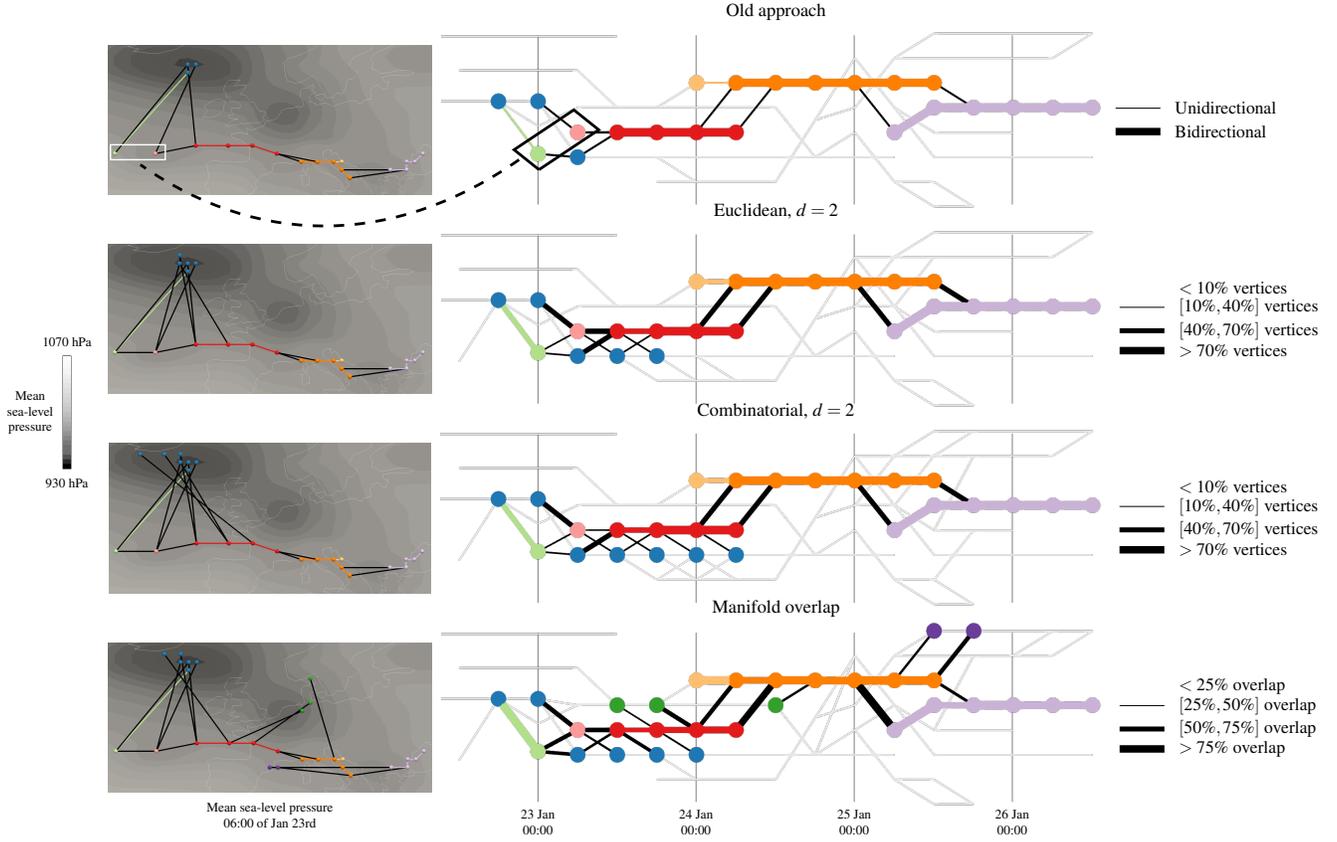}
    }
    \caption{The result of tracking minima across Europe between January 23rd - January 27th, where we have extracted the track spatially close or connected to the tracks representing the extratropical cyclone Klaus. In the top row, we show the result of the tracking using a previous approach \cite{Engelke2020}, while the three bottom rows show the tracks for our suggested approaches. They all manage to capture an edge connection that a previous approach completely misses. The nodes and edges in the graph, as well as the edges in the spatial embedding, are colored by extrema id. The correspondence probability is encoded in the line thickness of the edges of the tracking graph. In the visualization of the previous method, we encode the direction of the connection in the tracking graph as we receive only a binary mapping. All methods can capture the cyclone splitting at 06:00 on 24th January and the merging of it at 18:00 on 24th January, where it continues across the Mediterranean until it dies over the Black Sea.}
    \label{fig:klaus}
\end{figure*}

\section{Postprocessing of the tracking result}
\label{sec:post}

%\section{Postprocessing of the tracking result}
%\label{sec:post}

%\todoingrid{This is a very preliminary section that might even completely disappear in the final version}

While the main focus of the paper is on the generation of the probabilistic extrema correspondence, we want to briefly highlight some of the post-processing possibilities when applying the tracking in an application. Thereby we focus mostly on the aspect of simplification and building tracking hierarchies.
The input of this step is the output of the first part of our pipeline~(\autoref{sec:method}): sets of extrema $\mathcal{M}_{t}$, the extrema clustering hierarchy, and the overlap matrices  $\Overlap^\pm(t)$, for all time steps $t$.

\subsection{Hierarchical feature definition}
%\todoemma{The problem I see here is that I do in fact apply simplification of 0.5\% for the cyclone application case, only flat plateau artifacts for the other one but still}

As we only apply minimal simplification, in our examples $0.5\%$,  for the extrema extraction, the result can be very large tracking graphs of extrema.  So, it is important to be able to simplify and adjust the feature definitions to the applications in a post-processing step. Therefore, we will consider features $F_k \in \mathcal{F}_t$,  defined as hierarchical clusters of extrema of which crown features are an example~\cite{Nilsson2022Cyclone}. 
\begin{equation}
F_k =\{m_i \in \mathcal{M}_t|i\in I_k\}, 
\end{equation}
where $I_k \subseteq I_t$ is a set of indices defining the feature $F_k$.
We use the index set $I_t$ when we refer to all extrema (maxima or minima) in time step $t$.
The goal is to derive feature correspondence matrices $\FCmatrix$  for the features based on feature overlap matrices $\FOverlap$ derived from the overlap matrices $\Overlap$ for the individual extrema. 
In the following, we restrict the description to maxima and descending manifolds.

\subsubsection{Feature correspondence using sampling overlap}
Let $F_k=\{m_i \in \mathcal{M}_t|i\in I_k\}$ be a feature of time step $t$ and $F_l=\{m_j \in \mathcal{M}_{t+1}|j\in I_l\}$  feature in time step $t+1$. 
To define the sampling overlap of these features, we consider the samples from all extrema of the feature $F_k$ and determine how often they fall into the union of the descending manifolds of the extrema in $F_l$ which is given by the cardinality of the following set
\begin{equation}
\left(\bigcup\limits_{i\in I_k}\domain_\vertices(m_i) \right)\cap \left(\bigcup\limits_{j\in I_l} \dMan_{t\pm1}^{m_j}\right)
=
\bigcup\limits_{i\in I_k,j\in I_l}  (\domain_\vertices(m_i) \cap \dMan_{t\pm1}^{m_j})
\end{equation}
As all descending manifolds are disjoint this results in forward/backward overlap values for each pair of features, $F_l $ and $F_k$
%\todoemma{what does $p$ mean here?}
\begin{equation}
\FOverlap^\pm_{k,l}(t) = \sum_{i\in I_k,j\in I_l}  |(\domain_\vertices(m_i) \cap \dMan_{t\pm1}^{m_j})|= \sum_{i\in I_k, j\in I_l}\Overlap^\pm_{i,j}(t).
\end{equation}
To derive the forward and backward feature correspondence matrices we normalize the overlap matrix by the total number of samples.
\begin{equation}
\FCmatrix^+_{k,l}(t) =\frac{ \sum_{i\in I_k, j\in I_l}\Overlap^+_{i,j}(t)}{\sum_{i\in I_k}\domain_\vertices(m_i) }.
\end{equation}

\subsubsection{Feature correspondence derived from manifold overlap}
Let $F_k=\{m_i \in \mathcal{M}_t|i\in I_k\}$ be a feature of time step $t$ and $F_l=\{m_j \in \mathcal{M}_{t+1}|j\in I_l\}$  feature in time step $t+1$. 
In the manifold overlap case, the overlap of the features is defined by the overlap of the union of the participating descending manifolds
\begin{equation}
    \left(\bigcup\limits_{i\in I_k}\dMan_t^{m_i}\right) \cap \left(\bigcup\limits_{j\in I_l}\dMan_{t+1}^{m_j}\right)
 =
 \bigcup\limits_{i\in I_k,j\in I_l}\left(\dMan_t^{m_i} \cap\dMan_{t+1}^{m_j}\right)
\end{equation}
As all descending manifolds are disjoint
 the feature overlap between $F_k$ and $F_l$ is then given as
\begin{equation}
\FOverlap^+_{k,l}(t) = \sum_{i\in I_k, j\in I_l}|\dMan_t^{m_i} \cap \dMan_{t+1}^{m_j}|= \sum_{i\in I_k, j\in I_l}\Overlap^+_{i,j}(t).
\end{equation}
For the case of manifold overlap correspondence, the overlap is independent of the tracking direction which means $\FOverlap^+(t_i)= \FOverlap^-(t_i+1)^T$.

To derive the forward and backward feature correspondence matrices we still have to normalize the overlap matrix. Therefore we need the size of the union over all contributing descending manifolds. We do not explicitly store the size of the descending manifolds of the extrema, they can however be derived from the overlap matrix summing the all overlaps with the next time step $|\dMan_t^{m_i}|=\sum_{j\in I_t} \Overlap^+_{i,j}(t)$.
\begin{equation}
|\bigcup\limits_{i\in I_k}\dMan_t^{m_i}|
= \sum_{i\in I_k, j\in I_t}  \Overlap^+_{i,j}(t)
\end{equation}
The summation over $j$ considers all extrema in time step $t$.
Finally, the feature correspondence matrix is defined as
\begin{equation}
\FCmatrix^+_{k,l}(t) =\frac{ \sum_{i\in I_k, j\in I_l}\Overlap^+_{i,j}(t)}{\sum_{i\in I_k, j\in I_t}  \Overlap^+_{i,j}(t)}.
\end{equation}

\subsection{Tracking graph generation}
%\todoemma{Needs proofreading}
To generate the tracking graph extrema respective features from consecutive time steps are connected according to their correspondence matrices. Thereby the extrema ids are propagated in the track. Therefore, different usages of the correspondence matrices can be imagined.
%Besides simplifying the tracking graph by using hierarchical features, we also want to highlight that the tracking graph can further be simplified and filtered based on the derived tracking information from $\Cmatrix^\pm_{i,j}(t)$/$\FCmatrix^+_{l,k}(t)$. 

%Building the tracking graph the information from these matrices can be used in different ways.\\
\emph{Bidirectional connectivity:} Different combinations of the forward and backward correspondences can be applied.
We can restrict ourselves to bidirectional connections, e.g. where $\FCmatrix^+_{l,k}(t)$ and $\FCmatrix^-_{l,k}(t+1)$ are nonzero, or allow for a looser connection keeping a connection also if it is only non-zero in one temporal direction. 
In the case of binary gradient tracking the restriction to bidirectional connections doesn't allow for merges and splits on the extrema level, this is not the case for the proposed probabilistic strategies. 
 
\emph{Connectivity strength:} To derive a connectivity strength between features we provide the option of choosing the maximum, average, or minimum correspondence from the forward or backward tracking.  

\emph{Probability thresholding:} Furthermore, we can filter the connections by their probability $p$, and either demand a bidirectional probability or an any-directional probability to be fulfilled.

\emph{Semantic filtering:} In addition, it is possible to filter the correspondences based on a scalar field value and spatial positions 
%as the features are provided as output to the method given 
according to domain-specific constraints. Implausible connections based on spatial distance can also be removed. 

\section{Results}
\label{sec:results}
%% ============================================================================
%% 9_Results: Case Studies
%% ============================================================================

We demonstrate the usefulness of our two proposed strategies for two applications: tracking extratropical cyclones and analyzing the evolution of the valence electronic density in a molecular simulation. We show we can catch correspondences that the previous approaches miss in the cyclone application, as well as the ability to reason about the connectivity of the features in both applications. 

\subsection{Extratropical Cyclone Tracking}
Within climate research, understanding the formation and evolution of cyclones is an essential part of developing global climate models. Cyclones are low-pressure storms where high-speed rotational winds circle around a low-pressure center, however, there is no clear mathematical definition of cyclones. Furthermore, cyclones vary between geospatial regions and temporal seasons and can have multiple low-pressure centers. Extratropical cyclones (cyclones outside the tropics) can develop from shallow minima in the pressure field, which makes them difficult to track with conventional methods, resulting in methods tailored to specific storms.

In this application, we use our proposed strategies to track the storm Klaus over Europe between 21-27 January 2009. 
The comparative study of cyclone tracking methods~\cite{Neu2013} by the climate community, and the previous gradient-based tracking approach by Nilsson \etal~\cite{Nilsson2022Cyclone}, use the same storm and, supporting comparison with previous approaches. It is highlighted in these previous studies that the beginning of this storm is especially challenging to track.
For a detailed overview of the evolution of the cyclone, we refer to~\cite{Liberato2011klaus}. We use the same data as in previous studies: \cite{Dee2011era}, a re-analysis dataset produced by the European Centre for Medium-Range Weather Forecast, which is a simulation dataset constrained to observations. The dataset is a time series of 2D mean sea-level pressure fields with a spatial resolution of $360^{\circ}\times$ $180^{\circ}$, where the cell resolution is $1.5^{\circ}$. The temporal resolution of the dataset is 6 hours with sample points at 00:00, 06:00, 12:00, and 24:00.

In \autoref{fig:klaus}, we show the result of tracking Klaus using the binary gradient-based approach (top row) in comparison to our proposed strategies. The scalar field is first simplified using an \changed{arbitrarily chosen} persistence threshold of $0.5\%$ of the scalar function range, \changed{which removes topological noise}, and from the simplified field we extract minima and the ascending manifolds. We track the minima with the previous gradient-based approach, a Euclidean distance sampling with $d=2$, a combinatorial distance sampling with $d=2$, and the manifold overlap. We create correspondence matrices going both back- and forward in time. The tracking graphs on the right-hand side of the figure were generated from the correspondence matrices of maximum correspondence. 
In the spatial embedding, the points and edges are colored by extrema id, where black means that the edge connects different extrema id's. Also, note that the vertices in the tracking graph visualizations are colored accordingly.

The correspondence matrices with our proposed approaches contain many matches with a low probability. To avoid clutter in the tracking graph visualization and to simplify the analysis of the tracks, we remove low-probability connections from the tracking graphs. For the neighborhood-sampling approaches, 
%we sum the vertices in both temporal directions and divide with the maximum of possible vertex overlap, resulting in a normalized probability range between $[0, 1]$, where 
we keep connections with probability $p > 0.1$. For the manifold overlap, we keep connections where the overlap correspondence is $p > 0.25$. Within the visualization, we have filtered away or grayed out all tracks which have no correspondence with cyclone Klaus, while any connections and nodes have been kept.

Observing the spatial embedding of the result from the previous approach in \autoref{fig:klaus} (top left), we can see there is no connection between the two leftmost minima, however, the track that goes on from the pink one is actually the track that is Klaus, and we know from Liberato \etal~\cite{Liberato2011klaus} that Klaus was already developing at this point. Therefore, within the application context, these two minima should have a connection. For all three proposed methods, we get a connection between these minima, see the three bottom rows in \autoref{fig:klaus}. The probability of the connection varies depending on the method, where the maximum manifold overlap yields the highest probability.

\begin{figure}
    \centering
    % \subfloat[23 Jan 06:00]{
    %     \includegraphics[width=0.48\linewidth, trim={4cm 2cm 4cm 4cm},clip]{Figures/klaus/sf-manifolds-ridgecase-t0-rev.png}
    % }
    % \subfloat[23 Jan 12:00]{
    %     \includegraphics[width=0.48\linewidth, trim={4cm 2cm 4cm 4cm},clip]{Figures/klaus/sf-manifolds-ridgecase-t1-rev.png}
    % }
    % \\
    % \subfloat[23 Jan 06:00]{
    %     \includegraphics[width=0.48\linewidth, trim={4cm 2cm 4cm 4cm},clip]{Figures/klaus/sf-manifolds-ridgecase-t0-rev.png}
    % }
    % \subfloat[23 Jan 12:00]{
    %     \includegraphics[width=0.48\linewidth, trim={4cm 2cm 4cm 4cm},clip]{Figures/klaus/sf-manifolds-ridgecase-t1-rev.png}
    % }
    \adjustbox{width=\linewidth}{
          \input{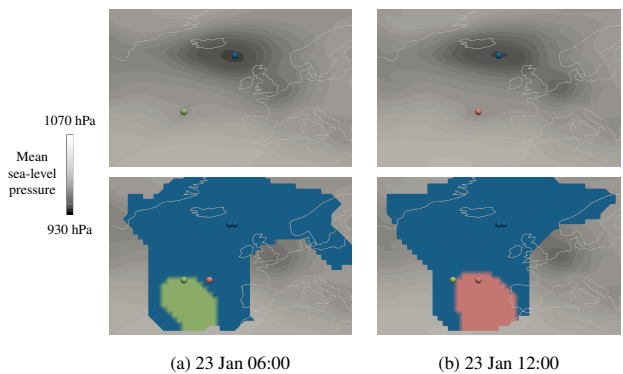}
    }
    \caption{\changed{In a) we show the mean sea-level pressure 06:00 23rd January on top, and on the bottom we show the embedding of the ascending manifolds for two minima of interest, and the corresponding minima in the next time step. In b), the mean-sea level pressure and manifolds of 12:00 23rd January are shown instead. The dark blue minima are deep minima within the pressure field with large ascending manifolds, while the light green and pink minima have smaller regions. In classic gradient-based tracking approaches, the light green and pink minima won't be matched as they are both within another minimum's ascending manifold, despite lying close to the boundaries of these manifolds.}}
    \label{fig:case-23jan}
\end{figure}

This part of the track is interesting because it fails both because of an extremum on \changed{a ridge in the scalar} field, and the existence of deep minima with a large basin over Iceland. In \autoref{fig:case-23jan}, we show the involved minima for the two time steps in both figures, while in a) we show the scalar field and ascending manifolds for 23rd January 06:00 and in b) for 23rd January 12:00. The darker blue minima originate close to Iceland, where we have a deep basin which attracts many integral lines and therefore a flat plateau in the simplified scalar field. We also see that the pink minimum lies close to the \changed{boundary} of the light \changed{green} minimum's ascending manifold in a), while the light \changed{green} lies exactly on the \changed{boundary between} the ascending manifolds in b). As mentioned earlier, see in \autoref{fig:klaus}, we manage to capture this case and calculate a probability for the correspondence, while the previous approach misses the connection entirely. However, note that because of the difficulty within this case, we get a lower probability for the connection between the light \changed{green} and pink minima than we do for the wrong connections to the dark blue minima, as the ascending manifolds take up such a large region in the scalar field.

\subsection{Electronic Density}

\begin{figure}[!ht]
    \centering
     \adjustbox{width=\linewidth, valign=b}{
          \input{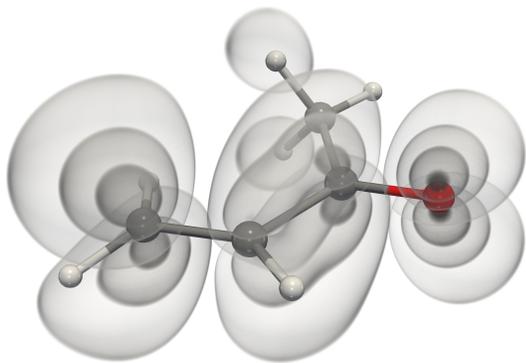}
        }
    \caption{A volume rendering of the electronic density distribution corresponding to the highest energy molecular orbital in butenone (\ce{C4H6O}) calculated at one of the steps during a molecular dynamics simulation. The molecule is shown in ball-and-stick representation for context, with red, gray, and white spheres representing Oxygen, Carbon, and Hydrogen atoms, respectively. Note that there are strong maxima on the opposite sides of some atoms in the molecule, see e.g. the red Oxygen atom, \changed{where a ridge in the scalar field with steep gradient on both sides separates the two charge density lobes}.}
    \label{fig:ed-molecule}
\end{figure}

\begin{figure*}[t]
    \centering
    \adjustbox{width=\linewidth}{
          \input{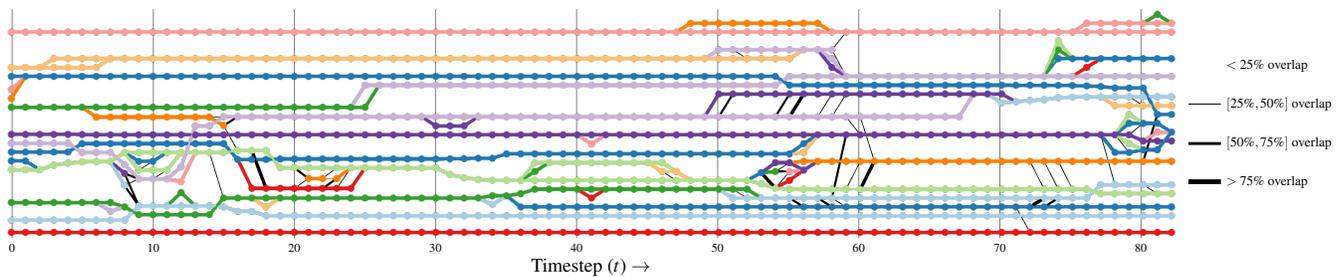}
    }
    \caption{An overview of the entire tracking graph of a molecular dynamics simulation, where the graph is generated based on the correspondences of maximum manifold overlap in both temporal directions, where we have selected only tracks with probability $> 0.25$. The nodes and edges are color-coded by the extremum index. The edge thickness encodes the probability of a correspondence.}
    \label{fig:ed-overview}
\end{figure*}

%\begin{itemize}
%    \item Show results with our different techniques (sampling within a distance from the extrema, overlapping manifolds)
%\end{itemize}

In the second case study, we consider the evolution of electronic density distribution in a molecular dynamics simulation.
More specifically, the subject of study, in this case, is the molecule called butenone (\ce{C4H6O}) along with a time-varying scalar field representing the valence electronic density distribution corresponding to the highest energy molecular orbital in this molecule. 
The high energy orbitals and the corresponding electronic density distribution are of extreme interest to the theoretical chemists as it captures the chemically active regions of the molecule and the molecular dynamic simulations augmented with quantum mechanical models allow the scientists to study how the spatial localization of electronic density changes and moves from one atom to other.
This data and task lend themselves naturally to topology-based approaches for extraction and tracking of key features in the density field.
However, in this case, one of the challenges is that the scalar field consists of steep ridges close to some of the maxima that may result in incorrect matches if traditional binary one-to-one gradient-based tracking approaches are applied. 
Therefore, the goal of this case study is to explore if the additional information provided by probabilistic gradient tracking results in better tracking compared to previous approaches.
\autoref{fig:ed-molecule} shows the volume rendering of the first time step in the simulation with the ball-and-stick model of the molecule shown for context. 
Note the two strong maxima around the red Oxygen atom, \changed{where a ridge in the scalar field with steep slopes are separating the two charge density lobes}. 

This molecular dynamics trajectory consists of 83 time steps. %in the simulation, where the first 62 scalar fields were sampled every 20th time step, and the next 10 every 4th and finally back to every 20th time step again. 
We first simplify the scalar fields to remove topological noise, after which the maxima of the electronic density field along with their descending manifolds are extracted. 
In this case study, we only derive probability based on the overlapping manifolds, as the proximity of the maxima around the atoms in the molecules would give connections across \changed{the ridges in the scalar field} which may not be meaningful within the application.

First, we show an overview of the entire tracks in \autoref{fig:ed-overview}, where the tracking graph has been generated based on the manifold overlap. 
We have taken the maximum of the overlap between two maxima to summarize the temporal directions and only kept correspondences with a probability greater than $0.25$. The edge thickness encodes the probability of the tracks, and we see that in general, the correspondences have a high probability. 
Some features persist through the entire simulation (e.g. the pink track on top), but many exist for tens of time steps. 
Yet, there are still some thin tracks in the graph, which could be close proximity features interacting. 
We observe that there are lots of merges and splits in the tracks between the time steps 45 to 65. 
This agrees with the domain scientists' expectations as well considering large geometric changes in the molecular configuration occur during this period.

We chose to inspect the simulation more closely between time steps 40 to 70. \autoref{fig:electronic_density} shows the results of our closer inspection, where we particularly analyze the split event between $t=[52, 57]$, and only highlight these tracks in the tracking graph in \changed{\autoref{fig:electronic_density}e}. The top row of \changed{\autoref{fig:electronic_density}a-d} shows spatial renderings of the electronic density field at a selection of time steps, where the spheres represent the maxima in question. 
In the closeups of the graph in \changed{\autoref{fig:electronic_density}f}, we show on top the results of the previous gradient tracking approach, where the edge thickness is now encoded as thin=unidirectional match and thick=bidirectional match. 
Within the split, we do not get any information on how probable the directions are, as the one-to-one mapping only supports the capture of the event, and we even miss some connections within the event.
In contrast, we encode the maximum manifold overlap in the bottom, where we can see that many of the maxima in the split actually are highly probable connections. 
It's only in the middle of the split that we have weak connections. Moreover, in the spatial embeddings in \changed{\autoref{fig:electronic_density}a-d}, we see that the structure itself slowly splits away, so that there are still connections to the light-green track which are not implausible in this application context.

\section{Conclusions and Future Work}
\label{sec:conclusions}
%% ============================================================================
%% 11_Conclusions: Conclusion
%% ============================================================================

We have proposed two strategies to encode the probability of correspondence between extrema when using gradient-based tracking: sampling the local neighborhood and calculating the overlap between the descending/ascending manifolds. As demonstrated in the first application example, all our proposed strategies can capture connections that a previous method could not. The generality of the approach supports the use of the proposed tracking in a wide variety of applications where extrema represent the features of interest, \changed{and as demonstrated, is useful even with minimal topological simplification with an arbitrarily chosen threshold. Furthermore, using our proposed approach with hierarchical features such as crown features \cite{Nilsson2020} should also help group unwanted extrema with the features in the application.} We also employ tracking both forward and backward in time for all strategies resulting in the tracks containing a large set of information that can be used to build a series of correspondence matrices based on multiple conditions. What conditions to apply is application-dependent, but without further refinement, the generated graphs will have a large edge-to-node ratio, resulting in graph visualizations that are cluttered and hard to analyze.

\changed{Both proposed strategies will keep the connections generated by the previous approach~\cite{Engelke2020, Nilsson2022Cyclone}, however, overlapping manifolds will in general result in more connections than sampling the neighborhood around the extrema. The overlapping manifolds is a global approach, taking all overlaps into consideration, while the sampling method is a local approach, given that the distance threshold is small compared to the size of the domain. Therefore, the results may vary between the different approaches, see a comparison in \autoref{fig:klaus}. While the sampling yields less cluttered tracking graphs, the optimal distance threshold to choose is not obvious before generating and analyzing the graphs.}

%especially in the case of overlapping manifolds. Sampling the local neighborhood with small distance thresholds yields less cluttered graphs, but the optimal distance threshold to choose is not obvious before generating and analyzing the graphs.

Another benefit of our proposed approach is that we can now reason about the correspondences with more information than before, as the previous approach could only tell where a connection was uni- or bidirectional. This is useful when analyzing merges and splits, as we see in the second application example that these are multi-time step events where small spurious features can appear that seem strongly connected, but in practice are irrelevant. \changed{However, the increase in information also increases the run time of the correspondence algorithms, as the ascending/descending manifolds have to be sampled multiple times. In particular, there is an overhead for the manifold overlap approach as the area of the manifolds is required for that strategy, which in the worst case necessitates calculation of areas of all manifolds in the scalar field. Furthermore, we currently use the number of vertices within a manifold to calculate the area/volume and manifold overlap, which is simple to use when the data is given on a regular grid. In future work, calculating the actual area/volume of the overlap would be preferred as that approach works for arbitrary 2/3D simplicial complexes.}

As our gradient-based tracking approach utilizes discrete gradients on a triangulated grid, the resulting tracks are still affected by flat plateaus and ridges in the \changed{scalar} field, where correspondences between extrema that are within these regions may be weaker. An interesting area of future work is improving the geometric embedding of the separatrices in the field, possibly using a combination of the discrete and numerical gradients when generating the descending/ascending manifolds, where the flat plateau problems could be alleviated. \changed{Furthermore, we also leave defining a distance function based on multiple fields for future work, which would be beneficial for applications with multivariate features, such as the cyclone tracking example in \autoref{sec:results}}.

\appendix
\section{Notations}
\label{sec:notations}
\begin{tabular}{cm{.49\linewidth}m{.33\linewidth}c}
\toprule
& Description & Notation\\
\midrule
& Simplicial domain  &  $\domain$ &\\
\cmidrule{2-3}
& Scalar field at time $t$  &  $f_t:\domain\rightarrow \R$ & \\
\cmidrule{2-3}
& Set of vertices  &  $\vertices$ & \\
\cmidrule{2-3}
& Vertices  &  $v_i, v_j \in \vertices$& \\
\cmidrule{2-3}
& Distance of vertices  &  $D(v _i,v_j)$& \\
\cmidrule{2-3}
& Distance threshold  &  $d$ & \\ 
\cmidrule{2-3}
& Extrema in one time step $t$  &  $\mathcal{M}_{t}=\{m_i|i\in I_t\}$ & \\
\cmidrule{2-3}
& Descending manifold of an extremum $m_i$ at time t  &  $\dMan_{t}^{m_i}$& \\
\cmidrule{2-3}
& Ascending manifold of an extremum $m_i$ at time t  &  $\aMan_{t}^{m_i}$& \\
\cmidrule{2-3}
& Cardinality of a set   &  $|\dots|$ & \\
\cmidrule{2-3}
& Set of sampling vertices  &  $\domain_\vertices$ & \\
\cmidrule{2-3}
& Overlap matrix for extrema comparing time $t$ with $t\pm 1$  &  $\Overlap^\pm_{i,j}(t)$& \\
\cmidrule{2-3}
& Overlap matrix for features comparing time $t$ with $t\pm 1$  &  $\FOverlap^\pm_{i,j}(t)$& \\
\cmidrule{2-3}
& Correspondence matrices for extrema comparing time $t$ with $t\pm 1$  &  $\Cmatrix^\pm_{i,j}(t)$ & \\
\cmidrule{2-3}
& Correspondence matrices for features comparing time $t$ with $t\pm 1$  &  $\FCmatrix^\pm_{i,j}(t)$ & \\
\cmidrule{2-3}
& Features as set of extrema  &  $F_k \! = \! \{m_i \in \mathcal{M}_t|i\in I_k\}$ & \\
\cmidrule{2-3}
& Set of features at time step $t$ &  $\mathcal{F}_t$ & \\
%$I_k\subset I_t$ is the respective index set of the extrema.\\
\bottomrule
\end{tabular}

\acknowledgments{
We thank Dr. Nanna H. List (KTH Royal Institute of Technology, Stockholm) for providing the dynamic electronic density simulation data that is used for one of the case studies in this paper. This work is supported by SeRC (Swedish e-Science Research Center), ELLIIT environment for strategic research in Sweden, and the Swedish Research Council (VR) grant 2019-05487.}

%\acknowledgments{
%The authors wish to thank A, B, and C. This work was supported in part by
%a grant from XYZ.}

%\bibliographystyle{abbrv}
\bibliographystyle{abbrv-doi}

\bibliography{ms}
\end{document}